# Prediction of three-dimensional flood-flow past bridge piers in a large-scale meandering river using convolutional neural networks

Zexia Zhang[1], Kevin Flora[1], Seokkoo Kang[2], Ajay B. Limaye[3], and Ali Khosronejad[1*]

[1] Civil Engineering Department, Stony Brook University, Stony Brook, NY 11794, USA

[2] Department of Civil and Environmental Eng., Hanyang University, Seoul 04763, South Korea

[3] Department of Environmental Sciences, University of Virginia, Charlottesville, VA 22904, USA

[*]Corresponding author, Email: ali.khosronejad@stonybrook.edu

## Abstract

The prediction of statistical properties of turbulent flow in large-scale rivers is important for river flow analysis. Large eddy simulations (LESs) provide a powerful tool for such predictions; however, they require a very long sampling time and significant computing power to calculate the turbulence statistics of riverine flows. In this study, we developed encoder–decoder convolutional neural networks (CNNs) to predict the first- and second-order turbulent statistics of the turbulent flow of large-scale meandering rivers. We trained the CNNs using a dataset obtained from the LES of the flood flow in a large-scale river with three bridge piers, which formed the training testbed. Subsequently, we employed the trained CNNs to predict the turbulent statistics of the flood flow in a river with different bridge pier arrangements, which formed the validation testbed. The CNN predictions for the validation testbed river flow were compared with the simulation results of a separately performed LES to evaluate the performance of the developed CNNs. The results showed that the trained CNNs can successfully produce turbulent statistics of the flood flow in large-scale rivers, such as that chosen for the validation testbed.

**Keywords:** Convolutional neural network, large scale rivers, flood flow predictions, large-eddy simulation

## 1 Introduction

High-fidelity numerical simulations are a crucial tool for studying flood-flow dynamics and sediment transport in natural waterways [1-11]. However, high-fidelity simulations of natural rivers require extensive computational resources, owing to the high Reynolds numbers of the flow, wide range of scales of vortical flow structures, and large size of natural rivers. Thus, most numerical simulations are performed using simplified turbulence closure models and relatively coarse computational grid systems. For instance, Olsen and Stokseth [12] modeled the flow in a river using Reynolds-averaged Navier-Stokes (RANS) turbulence closure. Stroesser et al. [13] incorporated vegetation into an RANS model to simulate a historic flood event within the reach of the Lower Rhine River. Fischer et al. [14] used an RANS-based model to study sediment transport and vegetation-stress interactions in the Danube River under flood conditions. Recently, high-fidelity numerical simulations were performed to capture the instantaneous characteristics of flow in natural waterways using more advanced turbulence closures, such as detached eddy simulation [15] and large eddy simulation (LES) [5,16]. These studies have proven highly valuable for the greater understanding of flow dynamics and sediment transport processes in natural waterways; however, their high computational cost has limited their practical application in the engineering community.

To address the computational cost of high-fidelity simulations, data-driven machine-learning methods have been recently applied to study the fluid dynamics of complex flows [17-19]. For example, Fenjan et al. [20] predicted the field of a curved open-channel flow using a multilayer perceptron. Kochkov et al. [21] developed a computational fluid dynamic (CFD) accelerator to enhance the efficiency of computations, and therefore, the simulation speed. In particular, efforts have been devoted to improve the efficiency of turbulence modeling by using machine learning algorithms [22-24]. For instance, Duraisamy et al. [25] studied the potential of machine learning to improve the accuracy of closure models for turbulent and transition flows. Tracey et al. [26] demonstrated potential machine-learning algorithms to enhance and/or replace conventional turbulence models, such as Spalart-Allmaras. By using the gene-expression programming algorithm, Zhao et al. [27] developed an explicit Reynolds-stress model directly implemented into RANS equations. Novati et al. [28] used multi-agent reinforcement learning to obtain more accurate turbulence models. In addition, machine-learning algorithms have been used to develop reduced-order models (ROMs) for flow-field prediction. For example, Mohan and Gaitonde [29] used an orthogonal decomposition and long short-term memory (LSTM) architecture to develop an ROM for turbulent flow control. Lui and Wolf [30] developed a flow-field predictive method using a deep feed-forward neural network.

Additionally, convolutional neural network (CNN)-based encoder–decoder algorithms have been proven effective for generating three-dimensional (3D) realizations of the flow field. For example, Guo et al. [32] used a CNN to construct a steady flow field around bluff bodies. They showed that the encoder–decoder CNN can quickly estimate the laminar flow field around bluff bodies. Their approach was recently enhanced by other researchers (e.g., Sekar et al. [33], Bhatnagar et al. [34], Ribeiro et al. [35], and Hasegawa et al. [36]). In addition, an encoder–decoder CNN has been used to produce super-resolution realization of flow fields. For instance, Deng et al. [37] used a super-resolution generative adversarial network (SRGAN) and enhanced SRGAN to augment the spatial resolution of the measured turbulent flows. By using a CNN and hybrid downsampled skip-connection/multiscale model, Fukami et al. [18, 38] generated high-resolution flow fields from low-resolution data. They also examined the performance of four machine-learning methods (i.e., multilayer perceptron, random forest, support vector regression, and extreme learning machine) in numerous regression problems for fluid flows [18]. Liu et al. [39] developed a static CNN and a novel multiple-temporal-path CNN to conduct super-resolution reconstruction of turbulent flows using direct numerical simulation. Finally, combined with LSTM, the encoder–decoder CNN exhibits satisfactory performance for predicting the time series of transient flow fields [40-43]. Despite their significant contributions, prior CNN-based machine learning studies have not focused on flood-flow modeling in large-scale meandering rivers. This gap currently limits the application of machine learning algorithms to inform flow modeling in natural waterways that commonly flow in sinuous paths over an irregular bed topography [44].

In this study, we aimed at developing encoder–decoder CNNs for generating 3D realizations of turbulent flood flow in large-scale rivers with wall-mounted bridge piers. In particular, we attempted to develop CNNs to predict the time-averaged flow field of large-scale meandering rivers. The LES method was used to produce the training dataset required to train the CNN algorithms. For this purpose, we conducted LES of the flood flow in a large-scale river (i.e., a training testbed). The instantaneous and time-averaged LES results of the flood flow in the training testbed were used to train the CNNs. The CNN training was performed with and without a physical constraint that enforced a divergence-free condition on the time-averaged flow field. The trained CNNs were subsequently used to predict the time-averaged flow field of a new testbed river, which formed a validation testbed. The CNN predictions for the flood flow of the validation testbed were compared against the simulation results of a separately performed LES in that river.

Based on the comparisons between the LES results and CNN predictions, the developed encoder–decoder CNN algorithm was found to hold great potential for predicting time-averaged flow fields of natural rivers, although it is several orders of magnitude less computationally expensive than high-fidelity numerical simulations.

The remainder of this paper is organized as follows. Section 2 presents the governing equations of the numerical model and a validation study to examine the model accuracy for open-channel flow predictions. Section 3 presents the computational details of the proposed model. In Section 4, we describe the encoder–decoder CNN algorithm, and in Section 5, the results and discussion are presented. Finally, we conclude the study's findings in Section 6.

## 2 Numerical model

### 2.1 Governing equations

The numerical model solves the instantaneous, incompressible, spatially filtered Navier–Stokes equations in curvilinear coordinates. The nondimensional form of the equations in compact tensor notation can be expressed as follows [45]:

$$J\frac{\partial U^j}{\partial \xi^j} = 0, \tag{1}$$

$$\frac{1}{J}\frac{\partial U^i}{\partial t} = \frac{\xi_l^i}{J}\left(-\frac{\partial}{\partial \xi^j}\left(U^j u_l\right) + \frac{1}{\rho}\frac{\partial}{\partial \xi^j}\left(\mu\frac{g^{jk}}{J}\frac{\partial u_l}{\partial \xi^k}\right) - \frac{1}{\rho}\frac{\partial}{\partial \xi^j}\left(\frac{\xi_l^j p}{J}\right) - \frac{1}{\rho}\frac{\partial \tau_{lj}}{\partial \xi^j}\right), \tag{2}$$

where $J = |\partial(\xi^1, \xi^2, \xi^3)/\partial(x_1, x_2, x_3)|$ is the Jacobian of the geometric transformation from Cartesian coordinates $\{x_i\}$ to generalized curvilinear coordinates $\{\xi^i\}$; $\xi_l^i = \partial\xi^i/\partial x_l$ indicates the transformation metrics; $u_i$ is the $i^{th}$ Cartesian velocity component; $U^i = (\xi_m^i/J)u_m$ is the contravariant volume flux; $g^{jk} = \xi_l^i \xi_l^k$ are the components of the contravariant metric tensor; $p$ is the pressure; $\rho$ is the fluid density; $\mu$ is the dynamic viscosity of the fluid; and $\tau_{ij}$ is the subgrid stress tensor for LES [46], which was modeled using the dynamic Smagorinsky subgrid scale model [47]. The governing equations are discretized in space on a hybrid staggered/nonstaggered grid arrangement using second-order accurate central differencing for the convective terms and second-order accurate, three-point central differencing for the divergence, pressure gradient, and viscous-like terms [48]. The time derivatives were discretized using a second-order backward differencing scheme [46]. The discrete flow equations were integrated in time using an efficient, second-order accurate fractional step methodology coupled with a Jacobian-free Newton–Krylov solver for the momentum equations and a GMRES solver enhanced with the multigrid method as a preconditioner for the Poisson equation.

The curvilinear immersed boundary (CURVIB) method is used to handle the complex geometry of the bathymetry of meandering rivers and cylindrical wall-mounted bridge piers (Section 5) [49]. In the context of the CURVIB method, the grid nodes in the computational domain are classified into three categories: background grid nodes in the fluid phase, external nodes located inside the solid objects (e.g., channel's side walls and bed and bridge piers), and immersed boundary (IB) nodes, which are fluid nodes immediately adjacent to the solid–fluid interfaces. The governing equations are solved over the background grid nodes, whereas all external nodes are blanked out of the computation. The boundary conditions are prescribed at the IB nodes using the wall-modeling approach within the CURVIB framework [50].

## 2.2 Model validation

To validate the flow solver of the numerical model, we simulated a turbulent open-channel flow in a 90° bend, which was experimentally studied by Abhari et al. [51]. The experiment was conducted in a rectangular flume of dimensions 18.6 m (length) × 0.6 m (width), × 0.7 m (depth). The inner and outer bend radii of the flume were 1.5 and 2.1 m, respectively. The mean-flow depth and discharge were 0.2 m and 0.03 $m^3 s^{-1}$, respectively, which resulted in a mean-flow velocity of 0.25 $m s^{-1}$ and a Reynolds number of $5 \times 10^4$. In addition, a programmable electromagnetic liquid velocimeter was used to measure the streamwise and spanwise components of the velocity field within the bend.

The computational domain, which has the same geometry as the experimental flume, was split into two zones. The first zone included a 2.4-m-long straight channel upstream of the bend that was discretized with 921 × 227 × 104 computational grid nodes in the streamwise, spanwise, and vertical directions, respectively. The second zone comprised the remaining channels, including the 90° bend, which was discretized with 3545 × 227 × 104 computational grid nodes in the streamwise, spanwise, and vertical directions, respectively. Both grids were stretched in vertical and spanwise directions so that the first node off the wall was located at $z^+$ of approximately 20. A precursor LES was performed in the first zone of the flume to obtain a fully developed turbulent flow field [1]. In the precursor LES, a periodic boundary condition was used in the streamwise direction. Then, the instantaneous fully developed turbulent velocity field at the outlet cross plane of zone I was saved and used to describe the inlet boundary condition of the second zone of the channel. At the outlet of the second zone, we employed the Neumann boundary condition for the velocity field and turbulence quantities. The free surface in both parts was treated as a rigid lid [52]. A nondimensional time step ($= \Delta t \frac{U_b}{H}$, where $\Delta t$ is the physical time step, $U_b$ is the mean-flow velocity, and $H$ is the mean-flow depth) of 0.002 was used, and LES was continued until a converged solution was obtained.

The velocity field was measured at cross-sections of 30°, 60°, and 90°. For each cross-section, five velocity profiles in the water column (equally spaced in the channel width) were measured and used for comparison with the LES results. Fig. 1 compares the measured and LES-computed profiles of the time-averaged velocity field in the water column. As shown, the LES results for the three velocity components agree well with the measurement results.

## 3 Computational details

To examine the potential of the encoder–decoder CNN for predicting time-averaged 3D turbulent flow in large-scale rivers, we considered a fully developed turbulent flow past wall-mounted bridge pier in a virtual large-scale meandering river as the testbed.

The testbed river was a single bend of a meandering river constructed to represent typical scales of natural meandering rivers (Fig. 2). It was 100-m wide, 3.3-m deep, and 2110-m long, and was generated using a common geometric model for the centerlines of meandering rivers [53,54]:

$$\theta(s) = \theta_0 \sin\left(\frac{2\pi s}{\lambda}\right) + \theta_0^3 (J_s \cos\left(\frac{6\pi s}{\lambda}\right) - J_f \sin\left(\frac{6\pi s}{\lambda}\right)), \quad (3)$$

where $\theta$ is the local direction of the channel centerline, $s$ is the position along the centerline, $\lambda$ is the bend wavelength, $\theta_0$ is the peak angular amplitude, $J_s$ is the skewness coefficient, and $J_f$ is the flatness coefficient. The meander bend wavelength was fixed at 12-channel widths, which is a typical value for meandering channels [55]. The testbed river in this study used the parameter combination of $\theta_0$ = 80°, $J_s$ = 0, and $J_f$ = 0, which generated symmetric meander bends. To focus on the effects of planform curvature on the flow field, the channel bed was set as flat.

We used a configuration of bridge piers as the training case of the CNN (Fig. 2a) and a different configuration for validating the trained CNN (Fig. 2b). In the training testbed river, three virtual cylindrical bridge piers with diameters of 2 m were installed 1055-m downstream of the inlet of the flow domain and spaced evenly in the channel width. The three piers are marked as $P_1$, $P_2$, and $P_3$ in Fig. 2a. The validation testbed river included four bridge piers with different arrangements. Specifically, $P_1$ and $P_3$ were kept at the same location as in the training testbed, whereas $P_2$ was shifted 25-m downstream and $P_4$ was shifted 25-m upstream; both shifts were performed parallel to the centerline of the virtual meander.

The background grid system of the testbed rivers was constructed to fit the curved geometry of the meander. The side walls, bridge piers, and flatbed of the channels were discretized using unstructured triangular grid systems and immersed into the background mesh using the CURVIB approach. The LES of the flood flow in the virtual testbed rivers was performed under the periodic boundary conditions in the streamwise direction, whereas the free surface of the rivers was described using rigid-lid assumptions. Details of the geometrical and hydrodynamic characteristics of the testbed rivers and computational grid systems are presented in Table 1. The LES was continued until the computed instantaneous flow field was statistically converged. Convergence was determined by monitoring the evolution of the total kinetic energy of the flow. Subsequently, we time-averaged the computed flow fields until both the first- and second-order turbulence statistics converged. The obtained instantaneous and time-averaged LES results were then used to train and validate the encoder–decoder CNN machine-learning algorithms.

## 4 Description of encoder–decoder CNN machine-learning algorithm

The CNN used in this study comprises two parts: (1) an encoder and (2) a decoder. The encoder part maps the input image into the feature space through the "learning process." The decoder part then reconstructs the output image using the learned features. This concept is illustrated in Fig. 3. The encoder part has two types of layers: a convolutional layer and a sampling layer. The convolutional layer is embedded with a nonlinear activation function and performs the following "convolution operation" [33]:

$$\boldsymbol{y}_i = \sigma(\boldsymbol{k_i} \otimes \boldsymbol{x} + b_i), \tag{4}$$

where $\boldsymbol{y}_i$ is the $i^{th}$ feature learned by the convolutional layer, $\sigma$ is the nonlinear activation function, $\boldsymbol{k_i}$ is the $i^{th}$ trainable convolutional filter, $\otimes$ is the convolution operator, $\boldsymbol{x}$ is the input, and $b_i$ is the $i^{th}$ bias. A diagram of the convolutional operation is shown in Fig. 4. The $i^{th}$ convolutional filter traverses the entire input matrix to produce the $i^{th}$ map of the output feature. Because the convolution operation reduces the size of the output feature map, zeros can be used to pad around the boundary of the input to control the output size. By using a downsampling process, we reduced the spatial resolution of the hidden layers to make the modeling affordable. The sampling layer could be a pooling or convolutional layer with a large movement step size, known as stride. Herein, we used the large stride method to minimize the computational cost and utilize the entire information of the input space.

Finally, the decoder part of the CNN was the reverse of the encoder part. It is necessary to increase the spatial resolution of the feature space and prepare it for the output. Therefore, we used a deconvolutional layer to perform the upsampling process.

## 5 Results and discussion

To train the CNN algorithms, we employed the instantaneous and time-averaged turbulent flow field of the training testbed river obtained from the LES. Subsequently, the trained CNN was used to produce 3D realizations of the time-averaged flow field of the testbed river. We trained the CNN such that it used an instantaneous flow field to generate a time-averaged flow field. The training dataset included the LES-computed (a) snapshots of the flow field from all grid layers of the training testbed river at 10 time instants and (b) time-averaged flow field of the training testbed river. The validation dataset, however, included the LES-computed (a) snapshot of the flow field at one time instant and (b) the time-averaged flow field. For each flow-field variable, including the time-averaged velocity components $\bar{u}$, $\bar{v}$, and $\bar{w}$ and secondary statistics $\overline{u'u'}$, $\overline{v'v'}$, $\overline{w'w'}$, $\overline{u'v'}$, $\overline{v'w'}$, and $\overline{u'w'}$, we trained a separate CNN. Furthermore, to assess the accuracy of the CNN predictions, we used two statistical error indices, namely the mean absolute error (MAE) and the mean absolute relative error (MARE), which can be defined as follows [56]:

$$MAE = \frac{\sum_{i=1}^{N}|\psi_{i(CNN)} - \psi_{i(LES)}|}{N}, \tag{5}$$

$$MARE = \frac{1}{N}\sum_{i=1}^{N}\frac{|\psi_{i(CNN)} - \psi_{i(LES)}|}{\psi_{i(LES)}}, \tag{6}$$

where $\psi_{i(CNN)}$ is the variable predicted using the CNN machine learning algorithm, $\psi_{i(LES)}$ is the value obtained using the LES model, and $N$ is the total number of samples (i.e., the total number of computational nodes required to discretize the flow domains of large-scale rivers). Finally, the computational cost of conducting LES for these testbeds was several orders of magnitude higher than that of the CNN. Specifically, the LES of the validation testbed required approximately 2500 CPU hours. In contrast, to generate the flow field of the river, the trained CNN required approximately 0.5 CPU hours. The CPU time required for a successful LES of natural rivers in this study was consistent with those of the others (e.g., [4-6, 13-16, 46]). For instance, Constantinescu et al. [15] reported that the LES of a meander bed with approximately 12 million computational grid nodes required approximately 15600 CPU hours; adjusted for the number of grid nodes in this study (~5.6 million), their computations would require approximately 7800 CPU hours. A drastic difference in the CPU time required for the LES and CNN predictions illustrates the advantages of the CNN machine-learning algorithms and their potential for real-time prediction of turbulent statistics in natural rivers.

Finally, in the training process of the encoder–decoder CNN algorithm, the CNN fine-tuned its connection weights by minimizing the "loss function," which represents the difference between the CNN's predictions and the LES results (i.e., the training dataset). The loss function, $L_f$, can be expressed as

$$L_f = \frac{\sum_{i=1}^{N}\left(\psi_{i(CNN)} - \psi_{i(LES)}\right)^2}{N}, \tag{7}$$

where $\psi_{i(CNN)}$ is the predicted entity at computational node $i$ using the CNN machine-learning algorithm, $\psi_{i(LES)}$ is the LES-computed entity at computational node $i$, and $N$ is the total number of samples (i.e., the total number of computational nodes required to discretize the flow domains of large-scale rivers). The so-trained CNN algorithm did not consider any flow dynamics-related constraint. In an attempt to include a physical constraint in the training process of the CNN, we enforced the divergence-free condition of the incompressible flow:

$$L_f = \lambda_1 \frac{\sum_{i=1}^{N}\left(\psi_{i(CNN)} - \psi_{i(LES)}\right)^2}{N} + \lambda_2 \frac{\sum_{i=1}^{N}\left(div_{i(CNN)} - div_{i(LES)}\right)^2}{N}, \tag{8}$$

$$div_i = \frac{\partial \overline{u_i}}{\partial x_i} + \frac{\partial \overline{v_i}}{\partial y_i} + \frac{\partial \overline{w_i}}{\partial z_i}, \tag{9}$$

where $\lambda_1$ (=10.0) and $\lambda_2$ (=1.0) are the weight parameters and $div_i$ is the divergence of the time-averaged velocity field at computational node $i$, and $\overline{u_i}$, $\overline{v_i}$, and $\overline{w_i}$ are the time-averaged velocity components in the streamwise, spanwise, and vertical directions, respectively. In the following text, we first present the prediction results of the CNN algorithm without the physical constraint (Subsections 5.1 and 5.2); subsequently, in Subsection 5.3, we present the results of the CNN algorithm predicted using the divergence-free constraint.

### 5.1 Prediction of time-averaged flow field: first-order turbulence statistics

We began by training the CNN algorithm using the instantaneous and time-averaged flow-field data obtained from the LES. For this purpose, we first conducted an LES of the training testbed river to produce a fully converged instantaneous velocity field of the flood flow. The convergence of the instantaneous results was checked using the time history of the total kinetic energy of the flow field. Then, we continued the LES while time-averaging the flow field of the river. Using the time-history analysis approach reported by Khosronejad et al. [1], we ensured that the time-averaged flow-field data converged. Three separate CNNs were trained for each of the three velocity components. The CNNs for the streamwise, spanwise, and vertical velocity components are denoted as $CNN_u$, $CNN_v$, and $CNN_w$, respectively. The training datasets for $CNN_u$, $CNN_v$, and $CNN_w$ included the 3D instantaneous and time-averaged velocity components $\bar{u}$, $\bar{v}$, and $\bar{w}$, respectively, obtained from the LES. As shown in Fig. 5, the signal input to the CNN is a fully converged instantaneous velocity component and the output signal is the time-averaged velocity component. During the training process, the dataset input to the CNN comprised 10 randomly selected fully converged instantaneous velocity components, whereas the target signal included 10 time-averaged velocity components. Because the inputs of the training datasets were taken from different time steps, the trained CNN was expected to be time-step-independent. Below, we present the sequence of actions performed to train the CNNs.

a) Conduct an LES of the training testbed to produce a statistically converged instantaneous flow field of the river.
b) Continue the LES to compute the time-averaged flow field of the river.
c) Randomly select 10 snapshots of the instantaneous velocity field and set them as the input signal arrays.
d) Set 10 time-averaged velocity fields as the target signal array.
e) Train the CNN by feeding the data of step (c) as the input and enforcing the data of step (d) as the output signal.

The trained CNNs were then used to predict the time-averaged first-order statistics of the validation testbed. The validation testbed flow-field data were not used in the training process of the CNNs, and as mentioned above, both the arrangement and number of bridge piers in the validation testbed river essentially differed from those of the training testbed. To predict the 3D realization of the time-averaged flow field of the validation testbed river, the CNNs required a snapshot of the flow field as the input signal. A snapshot was produced using a separately performed LES. By conducting the LES of the validation testbed river, we produced converged instantaneous and time-averaged flow-field results. While the snapshot of the LES results was used as input to the trained CNN, the time-averaged LES results were used to validate the CNN predictions for the time-averaged first-order statistics. Thus, to validate the CNN predictions for the time-averaged velocity field, the following sequence of actions was performed.

a) Conduct the LES of the validation testbed river to produce a statistically converged instantaneous flow field.
b) Continue the LES to compute the time-averaged flow field of the validation testbed river.
c) Select a single snapshot of the instantaneous velocity field and set it as the input signal to the previously trained CNNs.
d) Using the data from step (c) as the input signal, we ran the previously trained CNNs to predict the time-averaged velocity field of the validation testbed river.
e) Compare the predicted time-averaged flow field of the CNNs in step (d) with those of the LES in step (b).

Figs. 6–8 present the contours of the time-averaged flow field for the validation testbed river obtained using the CNN and LES at the free surface. In addition, CNN predictions were quantitatively compared with the LES results along the spanwise profiles around the bridge piers. The CNN predictions closely resembled the LES results, especially in the wake region of the bridge piers. The accuracy of the CNN predictions against the LES results can also be examined using the MAE and RMSE indices for all computational nodes in three dimensions (Table 2).

### 5.2 Prediction of time-averaged flow field: Reynolds stresses

In an LES, the second-order turbulence statistics are calculated by continuous time-averaging of the flow field until the turbulence characteristics converge statistically. For instance, the streamwise normal stress $\overline{u'u'}$ can be calculated as follows:

$$\overline{u'u'} = \frac{1}{n}\sum_{t_0}^{t_n}(u-\bar{u})(u-\bar{u}), \tag{10}$$

where $u$ and $\bar{u}$ are the instantaneous and time-averaged streamwise velocity components, $u'$ is the fluctuation in the streamwise velocity component, $t_0$ is the time-step at the beginning of the time-averaging, $t_n$ is the time-step at the end of time-averaging, and $n$ is the total number of time-averaging time steps.

To train a CNN for Reynolds stress prediction, first, the training dataset was required to be developed: the input and output signals of the CNN. The output signal of the training dataset included statistically converged values of the Reynolds stress obtained from the LES of the training testbed river. For example, the output signal of $\overline{u'u'}$ can be obtained as follows:

$$\overline{u'u'}_{output} = \frac{1}{n}\sum_{t_0}^{t_n}(u_{LES} - \bar{u}_{LES})(u_{LES} - \bar{u}_{LES}) \quad (11)$$

where the subscript "*LES*" represents the LES-computed values.

However, the input signals of the training dataset comprise a combination of CNN predictions and LES results. In other words, the input signals are essentially 10 snapshots of a nonconverged Reynolds stress field (e.g., the instantaneous field of $u'u'$):

$$u'u'_{input} = (u_{LES} - \bar{u}_{CNN})(u_{LES} - \bar{u}_{CNN}) \quad (12)$$

where $\bar{u}_{CNN}$ is the time-averaged streamwise velocity component predicted by $CNN_u$, as described in the previous section, and $u_{LES}$ is the instantaneous velocity component obtained from the LES results. The input signals include 10 nonconverged instantaneous arrays of Reynolds stress, whereas the output signal is a single array of fully converged LES-computed Reynolds stresses.

Owing to the nature of the turbulent river flow, the fully converged LES-computed Reynolds stresses have a highly heterogeneous distribution throughout the training testbed river; therefore, the values of second-order statistics differ by nearly two orders of magnitude between, for instance, the wake region of the bridge piers and near the river inlet. Such heterogeneity leads to an unsuccessful training process. To address this issue, we employed a preprocessing approach to render the distribution of second-order statistics more homogeneous. For this purpose, instead of using $u'u'_{input}$ and $\overline{u'u'}_{output}$, we used their cube roots (i.e., $\sqrt[3]{u'u'_{input}}$ and $\sqrt[3]{\overline{u'u'}_{output}}$, respectively) as the input and output signals, respectively, for training $CNN_{uu}$ (Fig. 9). Thus, the steps to train the CNN can be summarized as follows:

a) The time-averaged LES results of the training testbed river (Section 5.1) were used to extract the Reynolds stresses and set them as the output signal.
b) Ten snapshots of the instantaneous velocity field were randomly selected.
c) The CNN trained in Section 5.1 was used to predict the time-averaged velocity field of the training testbed river. Because the CNN was originally trained using the LES data of the trained testbed river, its predictions for this river were virtually the same as those for the LES.

f) The data obtained from (b) and (c) were used to construct 10 nonconverged instantaneous Reynolds stress fields, as described in Eqn. 12, which constitute the input signal of the training dataset.
g) The CNN was trained by feeding the data of (g) as input and enforcing the data of (a) as the output signal.

Once the CNNs were trained using the abovementioned approach, we used the trained CNNs to generate the Reynolds stress field of the validation testbed river. The trained CNNs used a nonconverged instantaneous Reynolds stress field as input to produce a fully converged Reynolds stress field at their output layer, as described in the following steps.

a) Conduct an LES of the validation testbed river to produce a statistically converged instantaneous flow field.
b) Continue the LES to compute the *nonconverged instantaneous* Reynolds stress field of the validation testbed river.
c) Select a single snapshot of the *nonconverged instantaneous* Reynolds stress field and set it as the input signal to the previously trained CNNs.
d) Using the data obtained from (c) as the input signal, run the previously trained CNNs to predict the fully converged Reynolds stress field of the validation testbed river as the output signal.

We used the fully converged LES data of the validation testbed river (as mentioned in Section 5.1) to obtain the converged Reynolds stress field of the validation testbed river. Comparing the predicted converged Reynolds stress field of the CNNs in (d) with the LES results, we attempted to validate the CNN predictions. Figs. 10–14 depict the LES results and CNN predictions for the contours of the main Reynolds stresses and the turbulent kinetic energy (*tke*) of the validation testbed river. The figures also depict the profiles of the Reynolds stresses and *tke* in the spanwise direction, allowing for a quantitative comparison between the CNN predictions and LES results predictions. Overall, the CNN predictions resemble the LES results fairly well, especially in the wake region where the CNNs seem to be capable of capturing the wake patterns of the bridge piers reasonably well. Finally, Table 3 presents the MAE and RMSE of the CNN relative to the LES results, and the CNN prediction errors are found to be less than 1%.

The time-averaging of the LES results of large-scale riverine systems is often a rather expensive task that requires considerable computing power. The proposed machine-learning approach is several orders of magnitude less expensive than conducting LES, and thus, can help to alleviate the costly process of obtaining the converged time-averaged solution of the turbulent river flows. Although the proposed approach eliminates the need to carry out time-averaging of the flow field, it still requires the converged instantaneous flow field data of the LES as an input signal. Given that the instantaneous flow field of LES can be acquired with relatively less computational effort, the proposed CNN algorithm can enable affordable high-fidelity modeling of large-sale rivers.

### 5.3 Prediction of flow field by enforcing the divergence-free constraint: first-order turbulence statistics

This section shows the reconstruction of the architecture of the CNN used in Subsection 5.1 to include a divergence-free constraint in the training process of the CNN. As the three velocity components at the adjacent nodes are required to calculate the divergence of the time-averaged velocity field (Eqn. 9), a new CNN architecture is required to simultaneously handle and process input arrays of the three velocity components. A schematic of the new CNN architecture is shown in Fig. 15, where the newly reconstructed CNN model has three input channels corresponding to the three velocity components. Additionally, we employed 3D convolutional layers to process the 3D flow field. However, to reduce the size of the CNN algorithm, the convolutional kernel was selected to be 2D, which only included the streamwise and spanwise velocity components (i.e., the two dominant velocity components). The steps required to train and validate the CNN algorithm were similar to those presented in Subsection 5.1.

To examine the level of divergence in the CNN predictions, Fig. 16 plots the contours of divergence of the time-averaged flow field at the free surface generated by the CNN and LES models, as well as the spanwise profiles of divergence. The LES results were mostly divergence-free, except for small regions very close to the bridge piers, where the velocity gradient was maximal. The LES divergence could be reduced by increasing the resolution of the computational grid system. In the LES results, the divergence-free condition was achieved by solving the Poisson equation, such that the $L_2$-norm of computational errors reached machine zero. Given the characteristics of the $L_2$-norm, even after the Poisson equation converged, there were some computational nodes at which the divergence level was not machine zero. Moreover, the CNN without the divergence-free constraint led to a high level of divergence throughout the river (Fig. 16a). However, the divergence-free constraint resulted in a relatively lower level of divergence in the river. Table 4 presents the mean absolute divergence of the CNN predictions ($\overline{|Div_{CNN}|}$) and LES results ($\overline{|Div_{LES}|}$) as well as their $L_2$-norm ($\|Div\|_2$), which are defined as follows:

$$\overline{|Div|} = \frac{\sum_{i=1}^{N}|div_i|}{N} \tag{13}$$

$$\|Div\|_2 = \sqrt{\frac{\sum_{i=1}^{N}\left(div_{i(CNN)} - div_{i(LES)}\right)^2}{N}} \tag{14}$$

As seen in the table, the divergence-free physical constraint resulted in over 50% and 75% reduction in the mean absolute divergence and its $L_2$-norm. However, it remains to be seen whether the divergence-free constraint can improve the flow-field predictions of the CNN.

Figs. 17–19 plot the contours of the predicted time-averaged velocity field using the LES model and the CNN algorithm with and without the physical constraint. In addition, the spanwise profiles of the predicted time-averaged velocity field are shown in order to quantitatively compare the LES and CNN results. The physics-constrained CNN predictions agreed better with the LES results than those of the CNN without the physical constraint. The MAE and RMSE of the CNN and LES model predictions for the 3D flow field are listed in Table 5. Here, considering the divergence-free physical constraint led to a reduction in the MARE of the CNN predictions by 20% and 16% for the time-averaged streamwise and spanwise velocity components, respectively.

## 6 Conclusion

The capabilities of the encoder–decoder CNN machine-learning algorithm for predicting the turbulence statistics of large-scale virtual rivers with wall-mounted bridge piers were examined. To train the encoder–decoder CNNs, we used the LES results of a training testbed river. Both instantaneous and time-averaged flow-field data were used to train the CNNs. Two training approaches were used, with and without physical constraints. In the training approach without physical constraints, the CNN was developed such that the error between the LES results and the CNN predictions was minimized. In training with the physical constraints, the CNN was trained by minimizing both the prediction error and divergence of the time-averaged flow field.

To evaluate the accuracy of the trained CNNs, we tested their performance in a validation testbed river in which the arrangement of the bridge piers was different from that of the training testbed river. A separate LES was conducted to obtain the instantaneous and time-averaged results of the flow in the validation testbed river. Using the LES results of the validation testbed river as the baseline, we compared the performance of the two CNN algorithms (i.e., with and without the physical constraints). The comparisons showed that both CNN algorithms can produce the time-averaged flow field of a large-scale river with reasonable accuracy. In addition, the use of the divergence-free constraint can improve the accuracy of the CNN predictions by approximately 20% and 16% for the time-averaged streamwise and spanwise velocity components, respectively.

Moreover, the prediction results showed that the encoder–decoder CNNs can successfully generate 3D realizations of the time-averaged flow field in large-scale rivers and around bridge piers. Specifically, the trained CNNs can capture the turbulent statistics of the wake flow on the lee side of the bridge piers. Compared with the LES results, the MAE and RMSE error indices of the CNN-predicted time-averaged flow field peaked in the wake of the bridge piers. This trend is consistent with the findings of Ti et al. [57], who reported that the maximum prediction errors of their artificial neural network occurred in the wake region of wall-mounted turbines, where the flow is more complicated.

The computational cost of the CNN approach for generating the time-averaged flow field of large-scale rivers was 150 CPU hours, which is at least one order of magnitude smaller than that of LES (i.e., ~2500 CPU hours). The CNNs required 0.5 CPU hours to complete, whereas the LES required approximately 2500 CPU hours. However, the developed CNNs used instantaneous LES results to produce a time-averaged flow field, and therefore, the CPU hours to compute the instantaneous LES results were added to the CNNs' CPU hours to obtain 150 CPU hours. Overall, the proposed CNN algorithms can enable an affordable and reliable prediction of the time-averaged flow field in large-scale rivers. In the future, we will test different data processing methods and add more physical constraints to the CNN algorithms to enhance the accuracy of turbulence statistics.

**Acknowledgments**

The LES code (doi: 10.5281/zenodo.4677354) and the machine learning data (doi: 10.5281/zenodo.4677315) are available on Zenodo repository. This work was supported by the National Science Foundation (grants EAR-0120914 and EAR-1823530). Computational resources were provided by the College of Engineering and Applied Science at Stony Brook University.

**Table 1:** Geometrical and hydrodynamic characteristics of the virtual testbed rivers used for the training and validation of the CNN. $H$, $B$, and $L$ are the mean-flow depth, width, and length of the meandering testbed rivers, respectively. $\boldsymbol{S}$ is the sinuosity of the meander; $U_b$ is the bulk velocity of the flood flow; $Re$ and $Fr$ are Reynolds and Froude numbers, respectively, both of which are calculated based on the mean-flow depth and the bulk velocity; $N_x$, $N_y$, and $N_z$ are the numbers of computational grid nodes in streamwise, spanwise, and vertical directions, respectively; $\Delta x$, $\Delta y$, and $\Delta z$ are the spatial resolutions in streamwise, spanwise, and vertical directions; $z^+$ is the vertical resolution in the wall unit; and $\Delta t$ is the temporal resolution.

| | | | |
|---|---|---|---|
| $H$ (m) | 3.3 | $N_x \times N_y \times N_z$ | $2201 \times 121 \times 21$ |
| $B$ (m) | 100 | $\Delta x$ (m) | 0.96 |
| $L$ (m) | 2110 | $\Delta y$ (m) | 0.83 |
| $\boldsymbol{S}$ | 1.76 | $\Delta z$ (m) | 0.17 |
| $U_b$ (m s$^{-1}$) | 2.04 | $z^+$ | 13000 |
| $Fr$ | 0.36 | $\Delta t$ (s) | 0.08 |
| $Re$ | $6.74 \times 10^7$ | | |

**Table 2:** Statistical error indices for CNN predictions versus LES results of time-averaged velocity components in the validation testbed river. MAE and MARE are the mean absolute relative error and mean absolute error of the CNN predictions, respectively. $\bar{u}$, $\bar{v}$, and $\bar{w}$ are the time-averaged streamwise, spanwise, and vertical velocity components, respectively. $U_b$ is the bulk velocity of the river (=2.04 m $s^{-1}$).

| | **MAE** | **MARE** |
|---|---|---|
| $\bar{u} / U_b$ | 0.019 | 0.047 |
| $\bar{v} / U_b$ | 0.026 | 0.247 |
| $\bar{w} / U_b$ | 0.004 | N/A |

**Table 3:** Statistical error indices of the CNN predictions for the Reynolds stress components relative to the LES simulation results in the validation testbed river. MAE is the mean absolute error and MARE is the mean absolute relative error.

| | **MAE** | **MARE** |
|---|---|---|
| $\overline{u'u'}$ | $4.97 \times 10^{-4}$ | 0.1430 |
| $\overline{v'v'}$ | $4.38 \times 10^{-4}$ | 0.1060 |
| $\overline{w'w'}$ | $1.28 \times 10^{-4}$ | - |
| $\overline{u'v'}$ | $2.53 \times 10^{-4}$ | 0.5493 |
| $\overline{v'w'}$ | $7.24 \times 10^{-5}$ | - |
| $\overline{u'w'}$ | $7.28 \times 10^{-5}$ | - |
| *tke* | $4.03 \times 10^{-4}$ | 0.0867 |

**Table 4:** Mean absolute divergence ($\overline{|Div|}$) and $L_2$-norm ($\|Div\|_2$) of the CNN predictions with and without the divergence-free constraint. CNN represents the CNN algorithm without physical constraint and $CNN_{phy}$ represents CNN algorithm with the divergence-free constraint.

| | $\overline{\|Div\|}$ | $\|\|Div\|\|_2$ |
|---|---|---|
| **$CNN_{phy}$** | 0.010 | 0.011 |
| **CNN** | 0.023 | 0.046 |

**Table 5:** Statistical error indices for velocity component predictions of the encoder–decoder CNN relative to the LES results for the validation case. MAE is the mean absolute error and MARE is the mean absolute relative error. CNN represents the CNN model without physical constraint and $CNN_{phy}$ represents CNN model with physical constraint.

| $\mathbf{CNN_{phy}}$ | MAE | MARE | CNN | MAE | MARE |
|---|---|---|---|---|---|
| $\bar{u} / U_b$ | 0.016 | 0.038 | $\bar{u} / U_b$ | 0.019 | 0.047 |
| $\bar{v} / U_b$ | 0.014 | 0.122 | $\bar{v} / U_b$ | 0.025 | 0.145 |
| $\bar{w} / U_b$ | 0.002 | - | $\bar{w} / U_b$ | 0.004 | - |

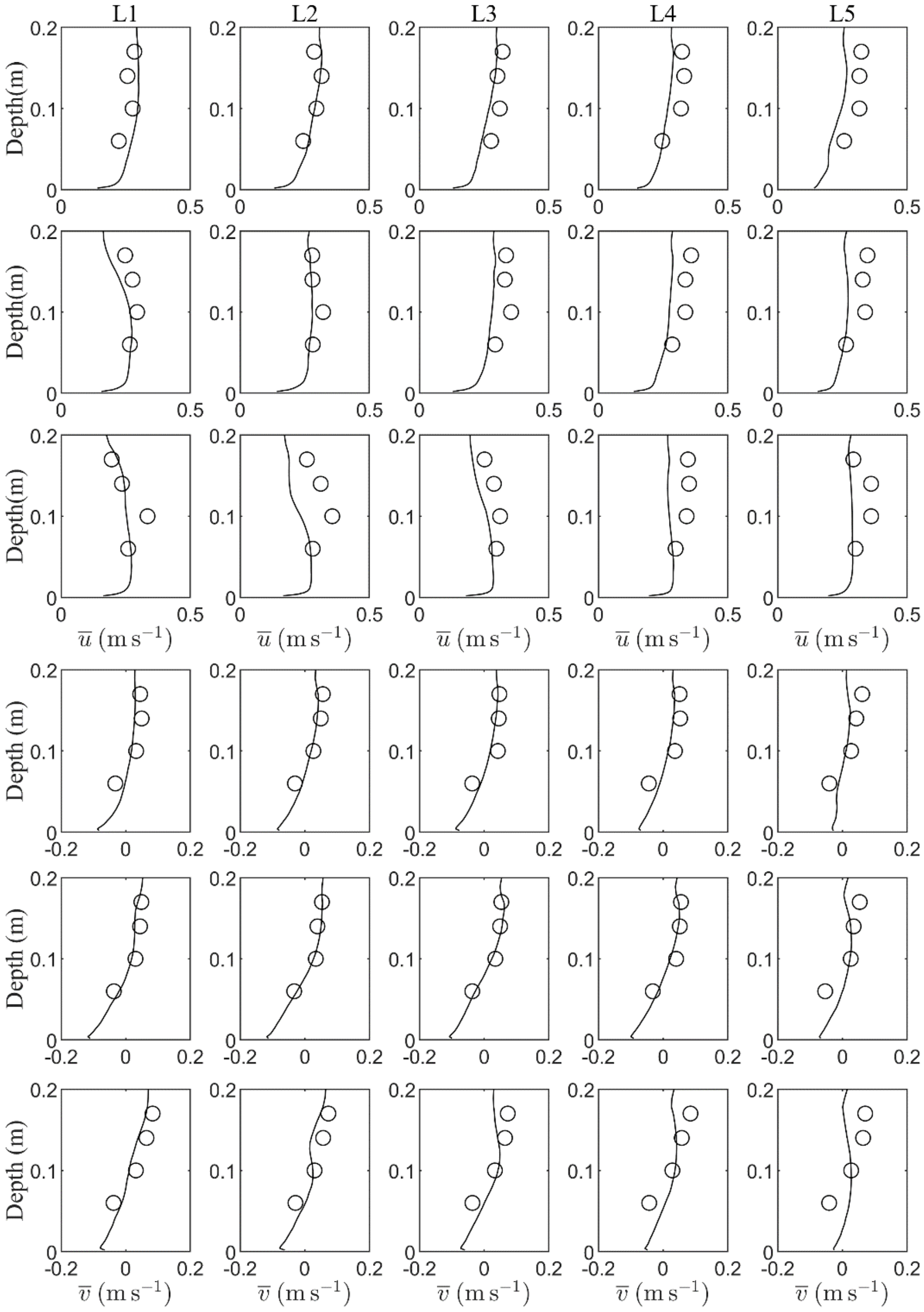

**Figure 1:** Measured (circles) and LES-computed (solid lines) profiles of the time-averaged streamwise ($\bar{u}$) and spanwise ($\bar{v}$) velocity components in vertical direction. $L_1$ to $L_5$ show the locations of profiles in spanwise direction. $L_1$ and $L_5$ are 0.1 m away from the inner and outer bends of the flume, respectively. $L_2$ to $L_4$ are located between $L_1$ and $L_5$, each 0.1 m apart.

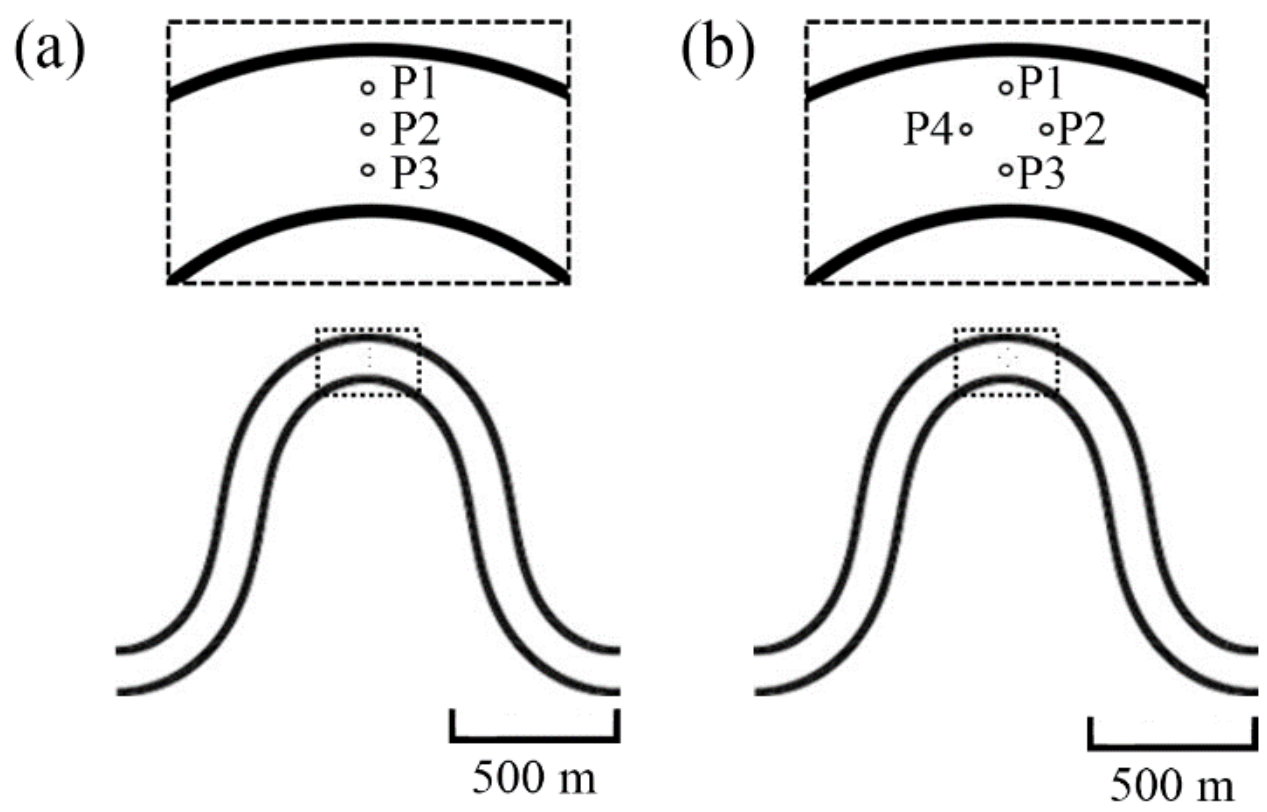

**Figure 2:** Schematic planforms of the virtual testbed rivers used for the training and validation of the CNN. The virtual cylindrical bridge piers are 2 m in diameter and installed 25 m apart at the apex of the river bend. Flow is from left to right.

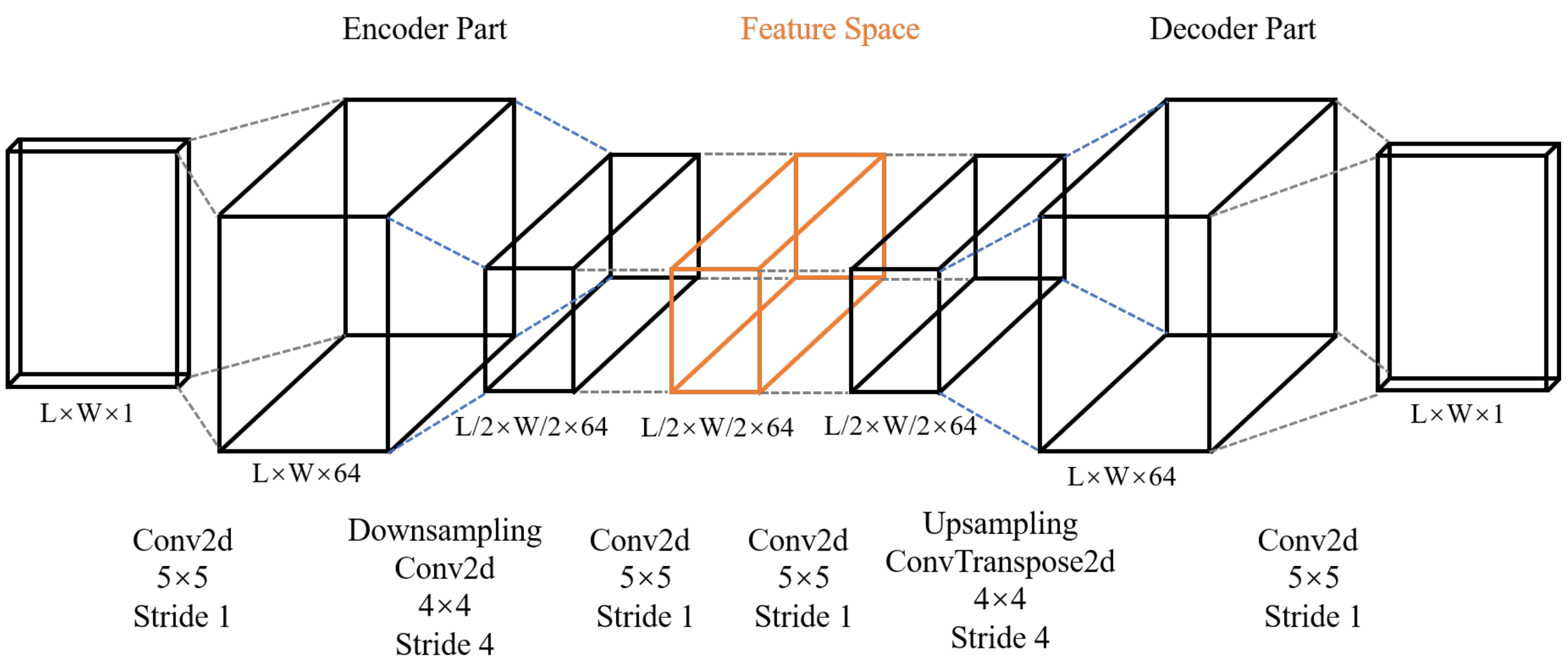

**Figure 3:** Schematic of the encoder–decoder CNN. Feature maps are depicted as solid boxes. Convolutional layers are depicted as gray dashed lines. Downsampling layer and upsampling layer are depicted as blue dashed lines. $L \times W \times$ channels represent the dimensions of each feature map. $L$ and $W$ represent the resolution of the input image in streamwise and spanwise directions, respectively. The layer type, kernel size, and stride size of each layer are shown below it. Strides represent the movement step-size of the convolutional filter.

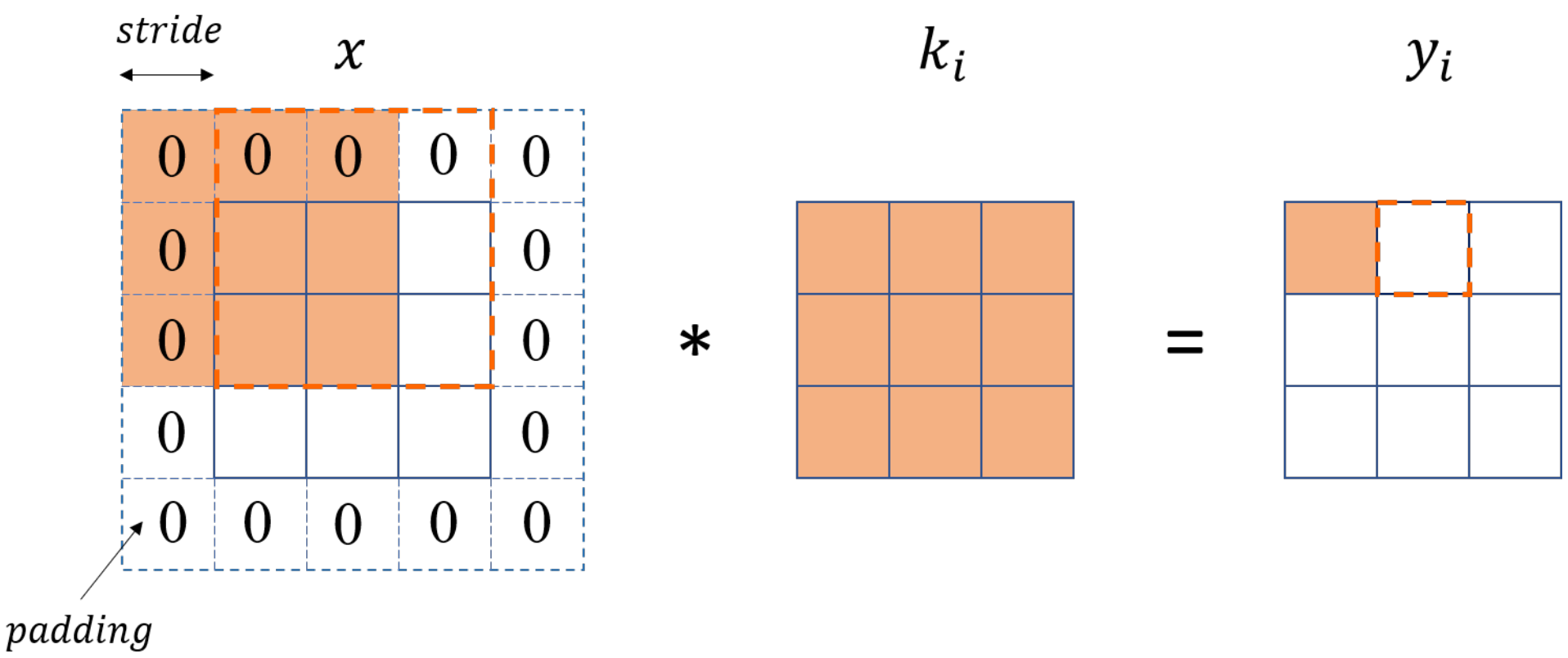

**Figure 4:** Schematic of convolution operation, where $x$ is the input, $k_i$ is the "$i^{th}$" convolution kernel, and $y$ is the output. The zeros around the input indicate padding. The orange squares represent input and output cells of the first convolution operation and the region around the orange dashed line show the cells present in the second convolution operation.

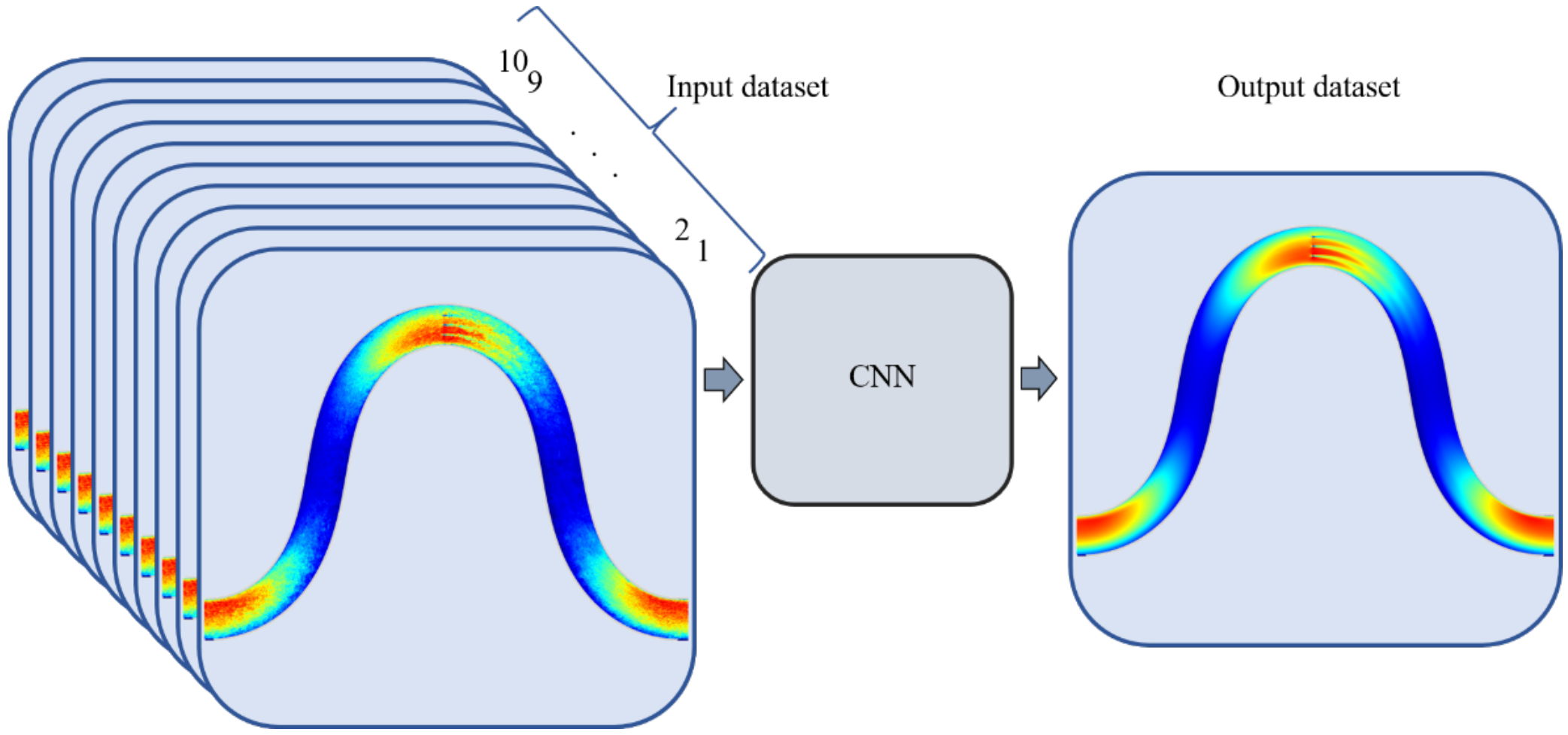

**Figure 5:** Schematic of the training procedure to develop $CNN_u$. The instantaneous stream velocity field is fed into $CNN_u$ as the input signal, whereas the time-averaged streamwise velocity field is enforced as the output signal. The input signals are obtained from 10 randomly selected instants, whereas the target signals are 10 same time-averaged results. Both instantaneous and time-averaged flow-field data were previously obtained from LES.

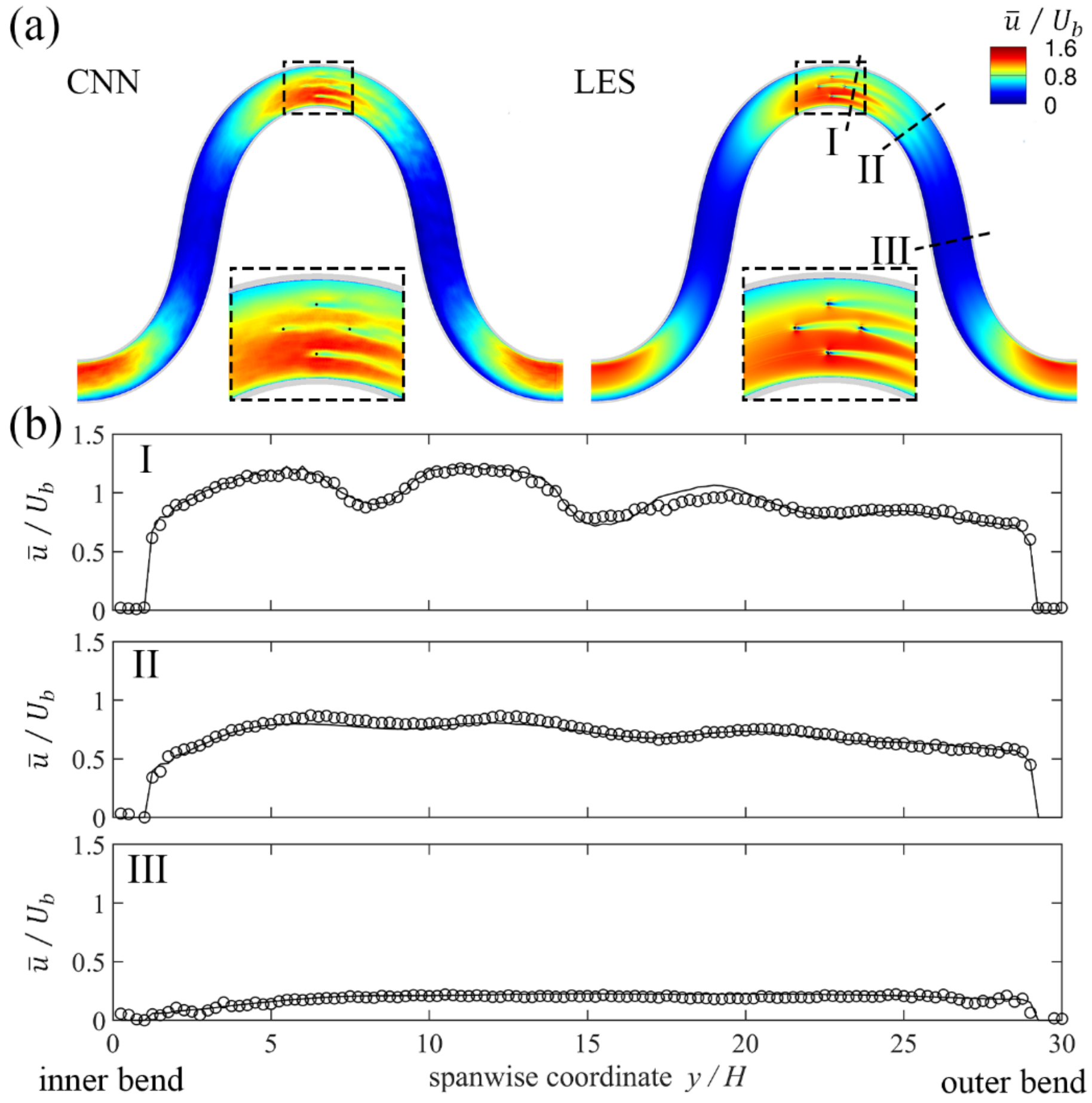

**Figure 6:** CNN predictions and LES results of the time-averaged streamwise velocity component for the validation testbed river. (a) Contours of $\bar{u}$ velocity component nondimensionalized with the bulk velocity ($\bar{u}$ / $U_b$) at the free surface of the river from the top view. In (a), flow is from left to right. (b) Profiles of $\bar{u}$ / $U_b$ in the spanwise direction along the three dashed lines of I, II, and III in (a). In (b), solid lines and circles represent the LES and CNN results.

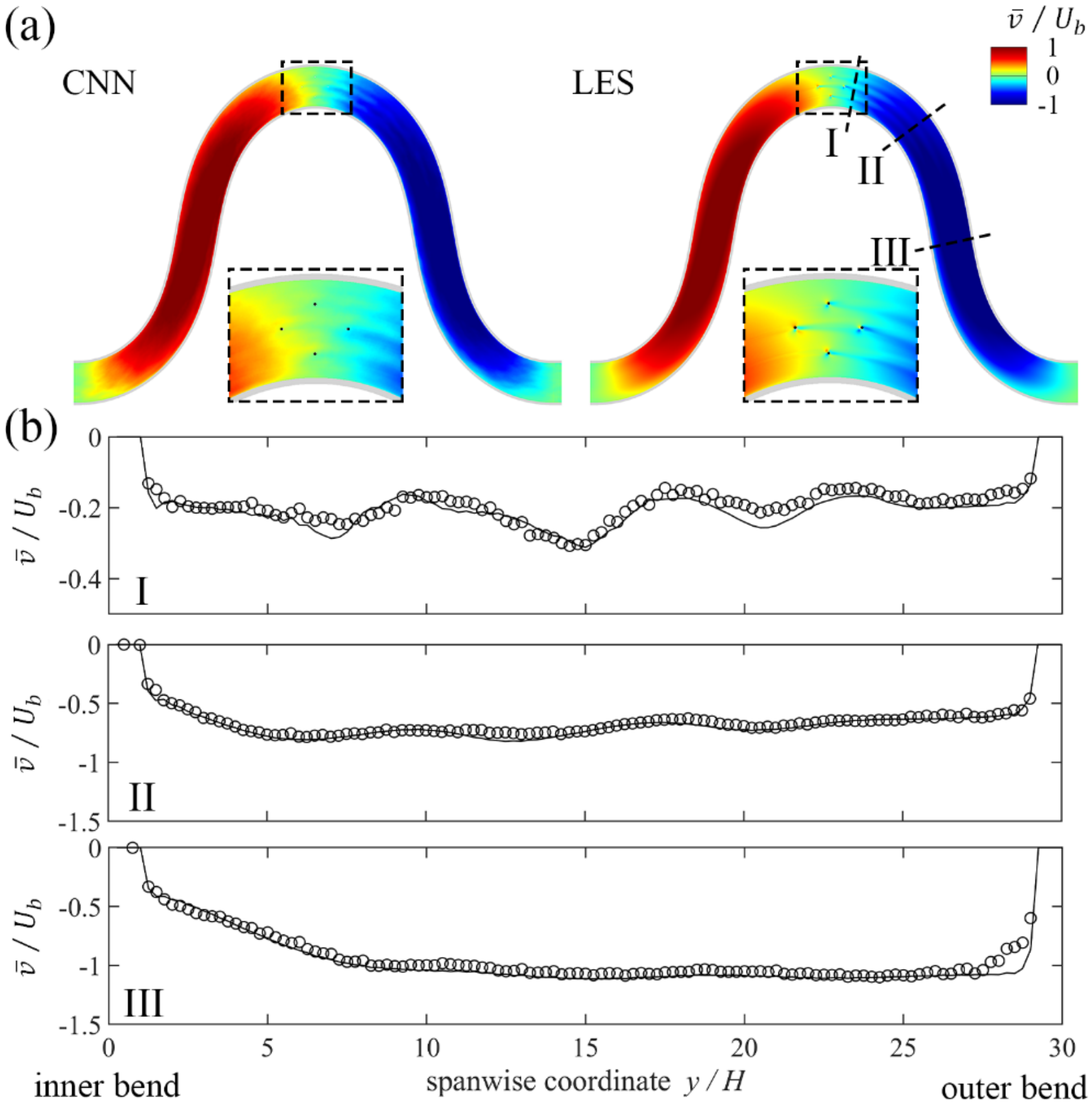

**Figure 7:** CNN prediction and LES results of the time-averaged streamwise velocity component for the validation testbed river. (a) Contours of spanwise velocity component nondimensionalized with the bulk velocity ($\bar{v}$ / $U_b$) at the free surface of the river from the top view. In (a), flow is from left to right. (b) Profiles of $\bar{v}$ / $U_b$ in the spanwise direction along the three dashed lines of I, II, and III in (a). In (b), solid lines and circles represent the LES and CNN results.

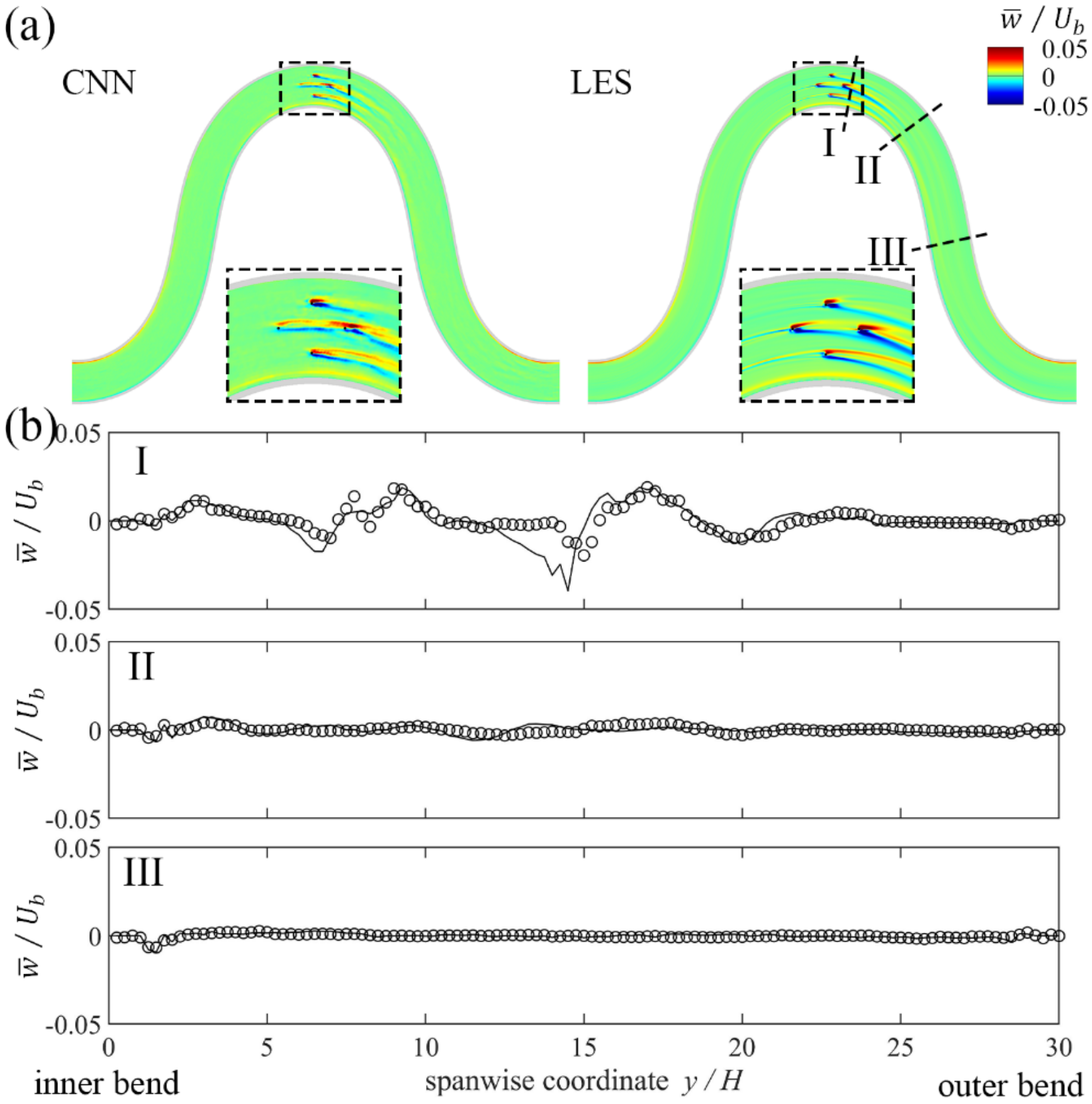

**Figure 8:** CNN predictions and LES results of the time-averaged vertical velocity component for the validation testbed river. (a) Contours of spanwise velocity component nondimensionalized with the bulk velocity ($\overline{w}$ / $U_b$) at the free surface of the river from the top view. In (a), flow is from left to right. (b) Profiles of $\overline{w}$ / $U_b$ in the spanwise direction along the three dashed lines of I, II, and III in (a). In (b), solid lines and circles represent the LES and CNN results.

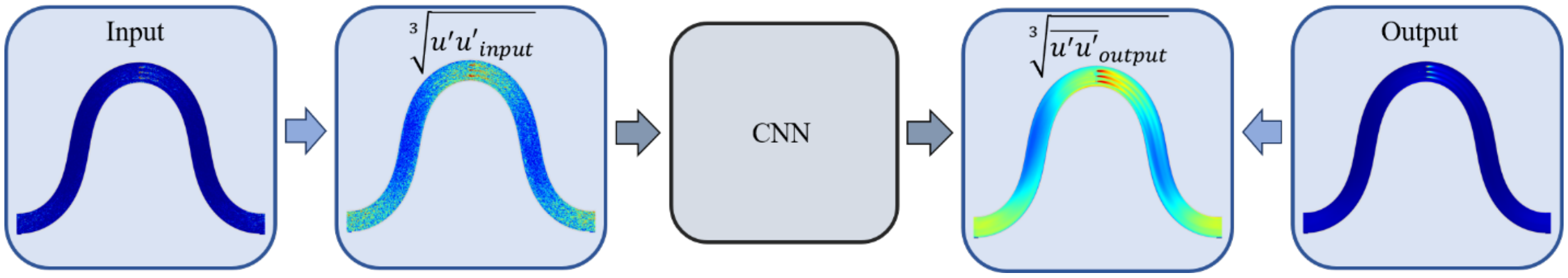

**Figure 9:** Schematics of preprocessing the input and output signals for CNN training. This is done by calculating the cubic root of the Reynolds stress values to render their distribution more homogenous.

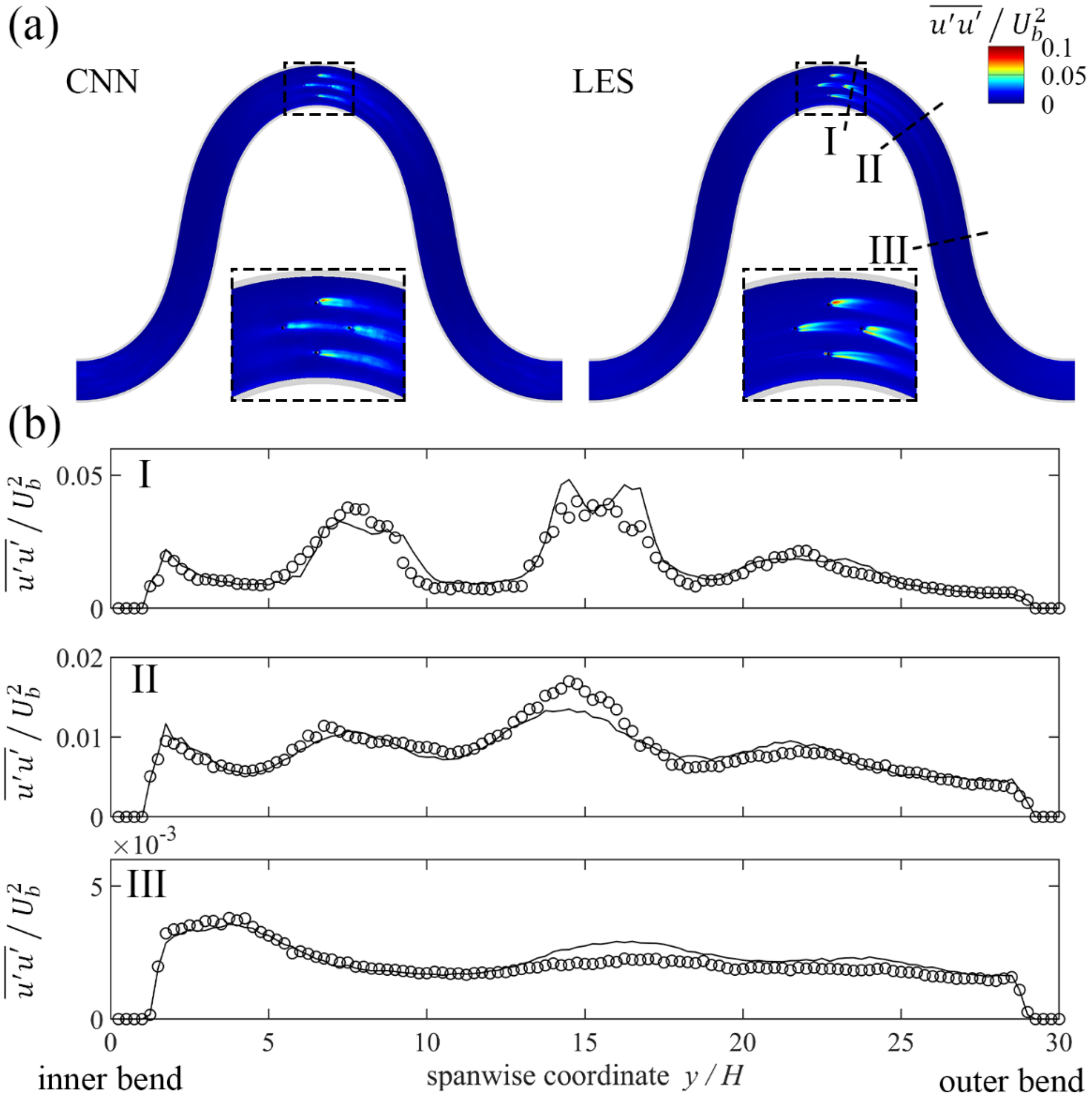

**Figure 10:** CNN predictions and LES simulation results of the streamwise normal Reynolds stress in the validation testbed river. (a) Contours of the Reynolds stress $\overline{u'u'}$, nondimensionalized with the square of bulk velocity ($U_b^2$), at the free surface of the virtual river from the top view and the flow is from left to right. (b) Profiles of the dimensionless Reynolds stress $\overline{u'u'}$ in the spanwise direction along the three dashed lines I, II, and III, as shown in (a). In (b), solid lines and hollow circles mark the LES and CNN results, respectively.

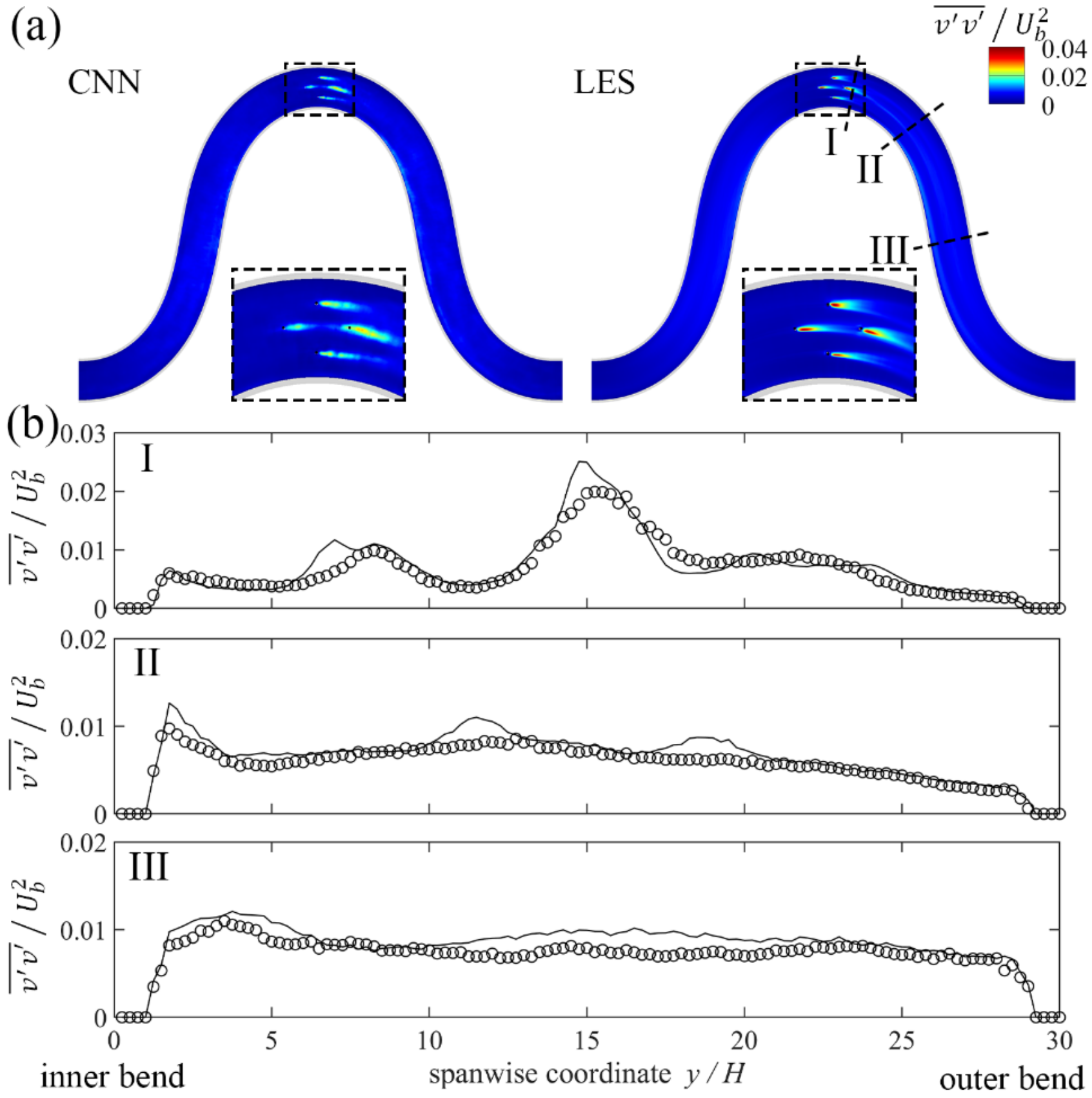

**Figure 11:** CNN predictions and LES simulation results of the spanwise normal Reynolds stress in the validation testbed river. (a) Contours of the Reynolds stress, $\overline{v'v'}$, nondimensionalized with the square of bulk velocity ($U_b^2$), at the free surface of the virtual river from the top view and the flow is from left to right. (b) Profiles of the dimensionless Reynolds stress $\overline{v'v'}$ in the spanwise direction along the three dashed lines I, II, and III, as shown in (a). In (b), solid lines and hollow circles mark the LES and CNN results, respectively.

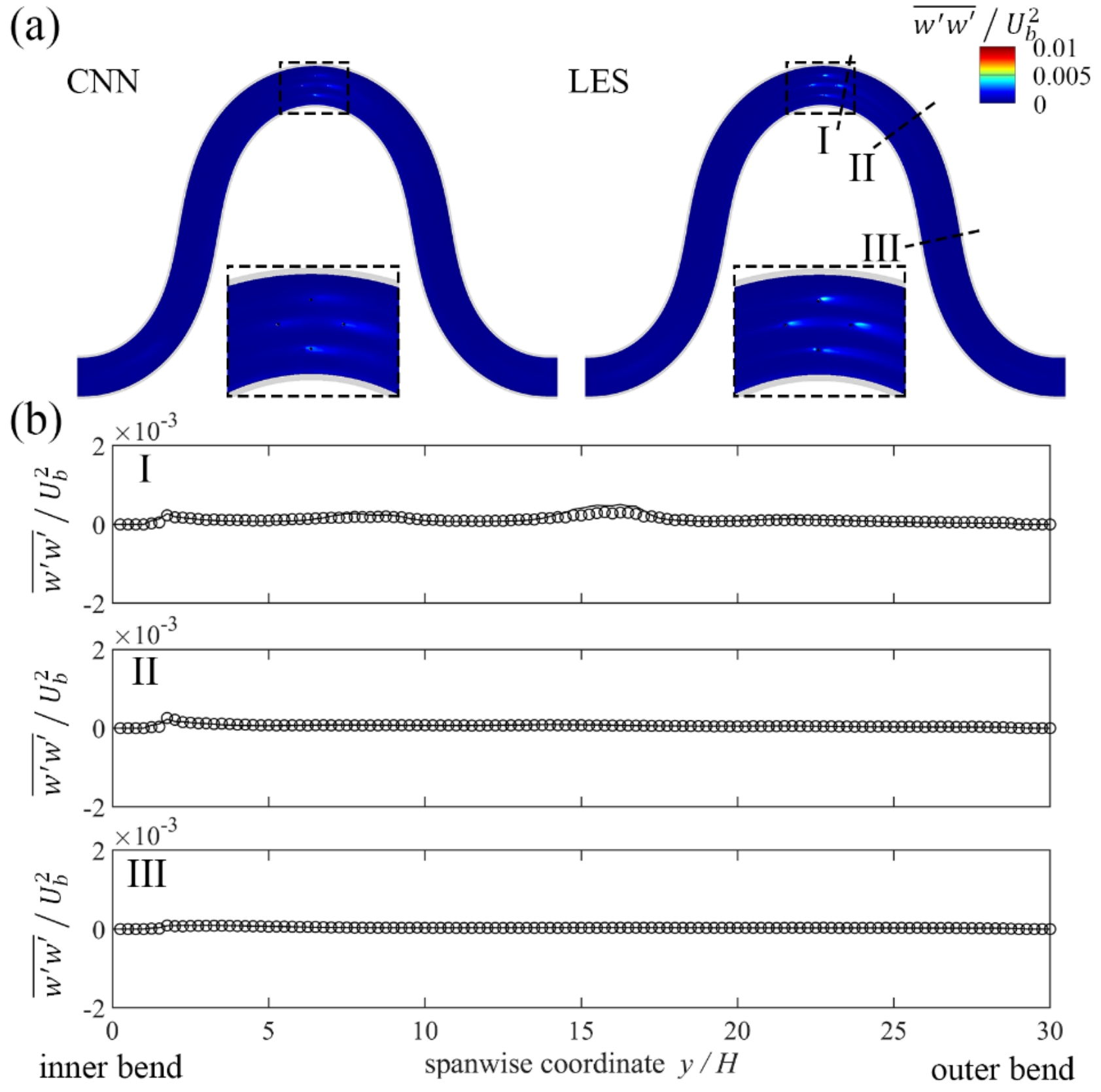

**Figure 12:** CNN predictions and LES simulation results of the vertical normal Reynolds stress in the validation testbed river. (a) Contours of the Reynolds stress, $\overline{w'w'}$, nondimensionalized with the square of bulk velocity ($U_b^2$), at the free surface of the virtual river from the top view and the flow is from left to right. (b) Profiles of the dimensionless Reynolds stress $\overline{w'w'}$ in the spanwise direction along the three dashed lines I, II, and III, as shown in (a). In (b), solid lines and hollow circles mark the LES and CNN results, respectively.

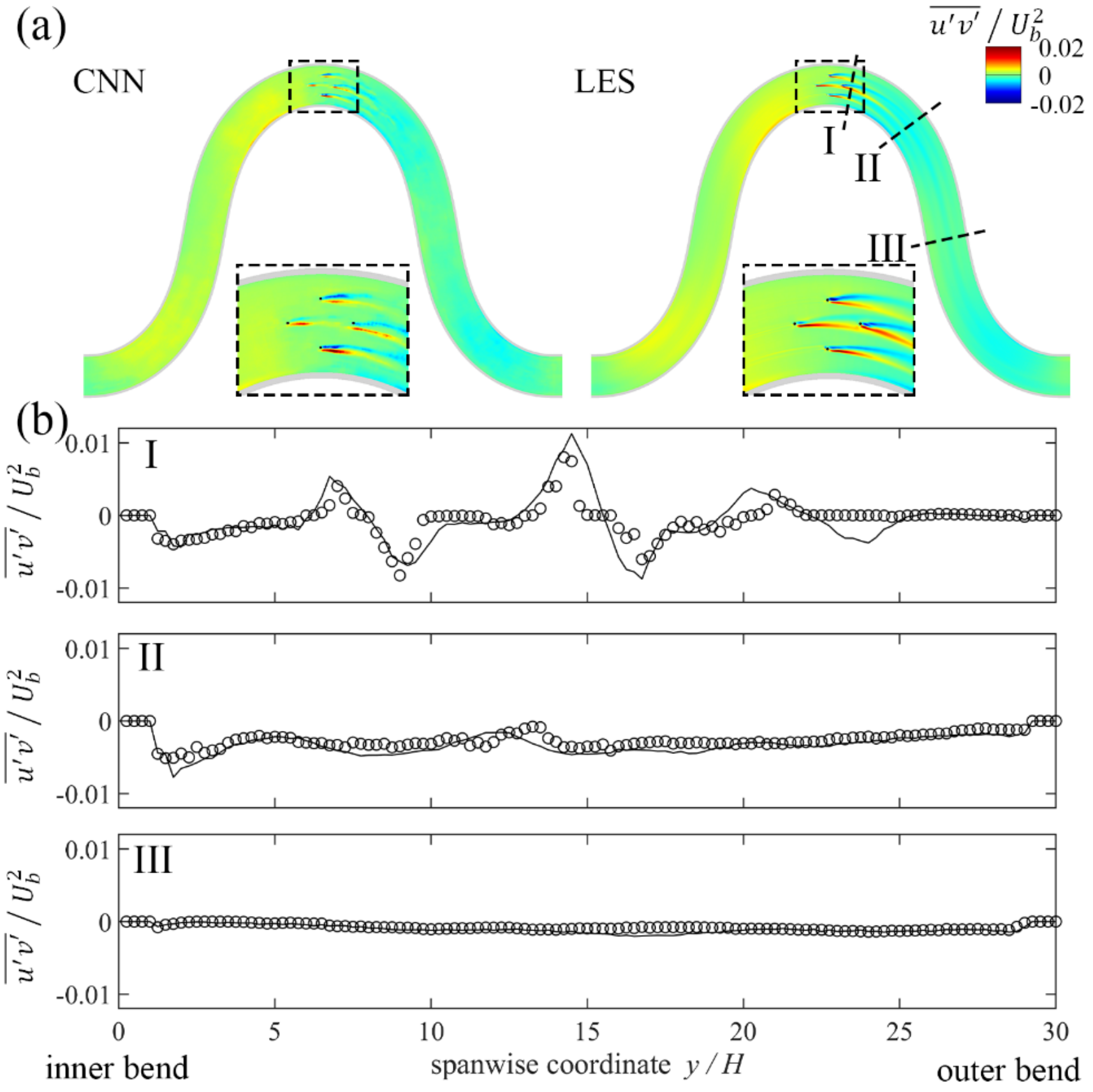

**Figure 13:** CNN predictions and LES simulation results of the Reynolds stress $\overline{u'v'}$ in the validation testbed river. (a) Contours of the Reynolds stress, $\overline{u'v'}$, nondimensionalized with the square of bulk velocity ($U_b^2$), at the free surface of the virtual river from the top view and the flow is from left to right. (b) Profiles of the dimensionless Reynolds stress $\overline{u'v'}$ in the spanwise direction along the three dashed lines I, II, and III, as shown in (a). In (b), solid lines and hollow circles indicate the LES and CNN results, respectively.

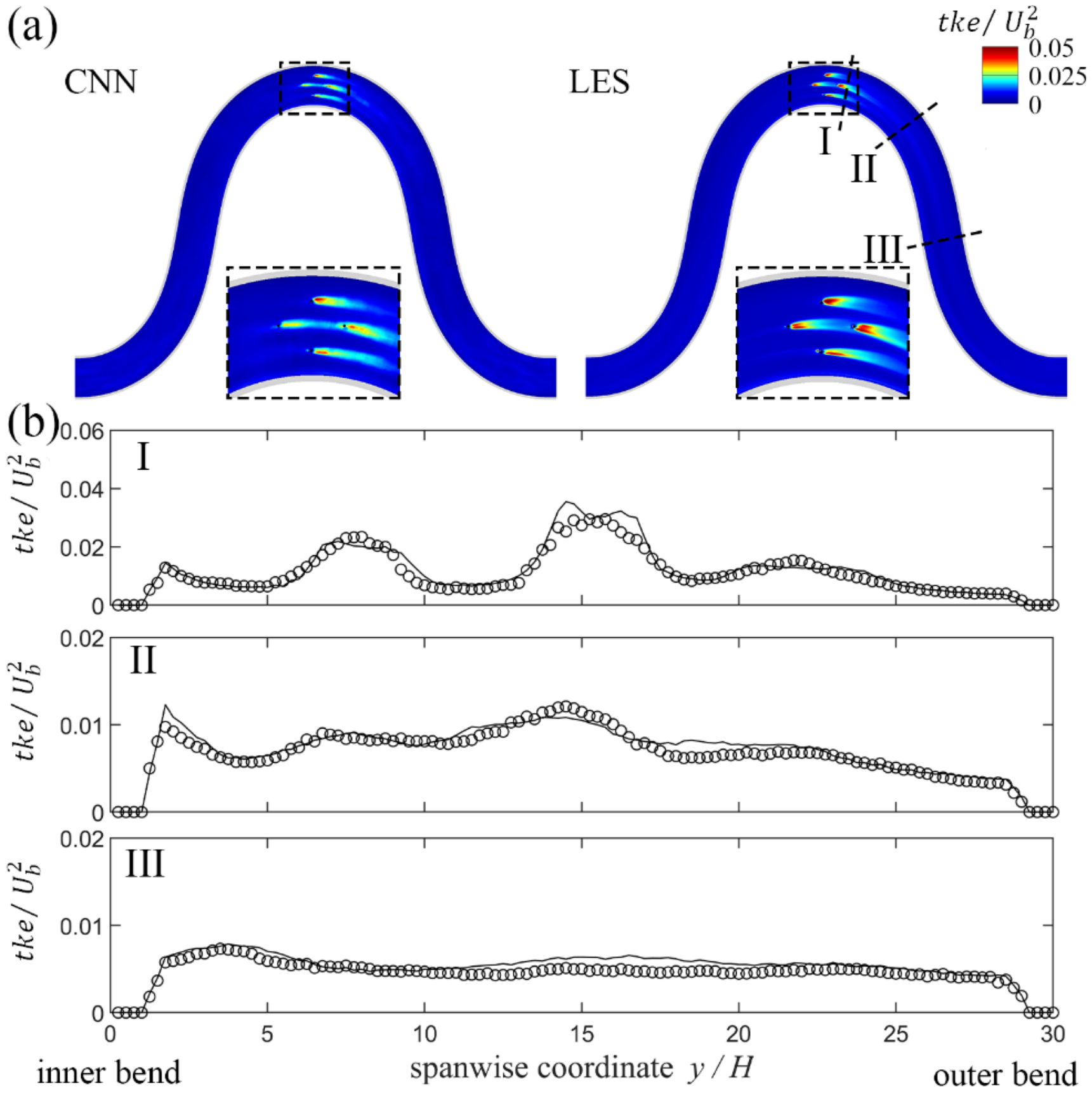

**Figure 14:** CNN predictions and LES simulation results of the turbulent kinetic energy (*tke*) in the validation testbed river. (a) Contours of the *tke*, nondimensionalized with the square of bulk velocity ($U_b^2$), at the free surface of the virtual river from the top view and the flow is from left to right. (b) Profiles of the dimensionless *tke* in the spanwise direction along the three dashed lines I, II, and III, as shown in (a). In (b), solid lines and hollow circles mark the LES and CNN results, respectively.

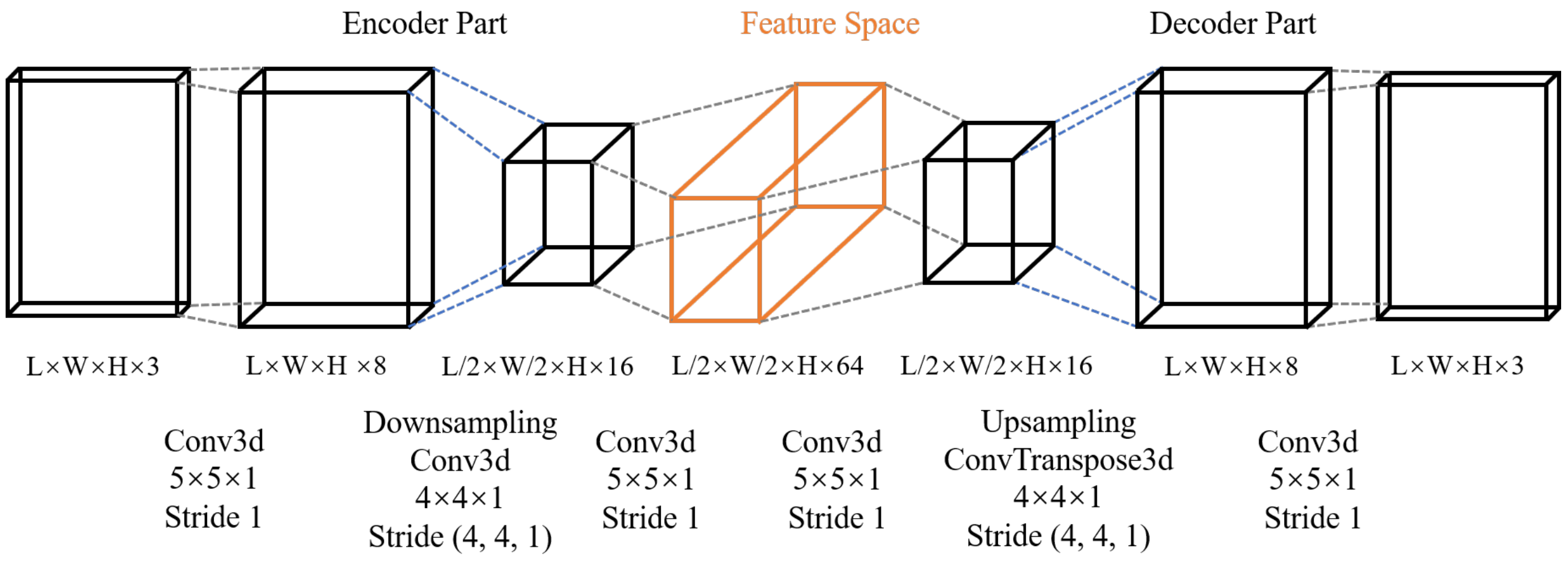

**Figure 15:** Scheme of the encoder–decoder CNN designed to accommodate the divergence-free constraint during the training process. The feature maps are shown as solid boxes. The convolutional layers are depicted as dashed gray lines. The downsampling and upsampling layers are depicted as blue dashed lines. The $L \times W \times H \times channels$ represent the dimensions of each feature map, where $L$, $W$, and $H$ represent the resolution of the input image in the length, width, and height directions, respectively. The layer type, kernel size, and stride size are shown below each layer.

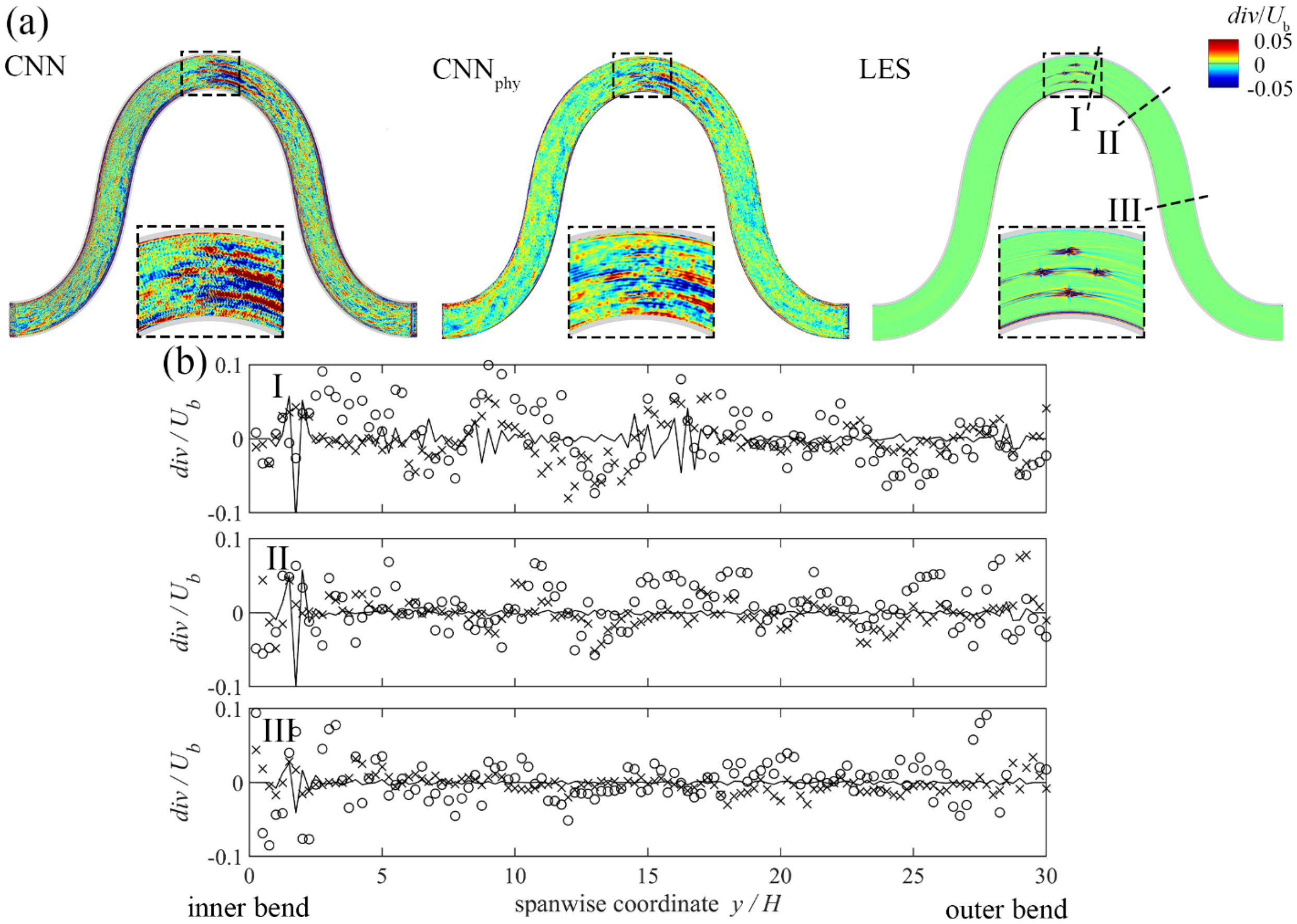

**Figure 16:** Time-averaged CNN predictions and LES results obtained for the 3D flow field of the validation testbed river. Contours of divergence ($div/U_b$) at the free surface of the river are shown from the top view. CNN: CNN predictions without the divergence-free physical constraint, $CNN_{phy}$: CNN predictions with the physical constraint. Flow is from left to right.

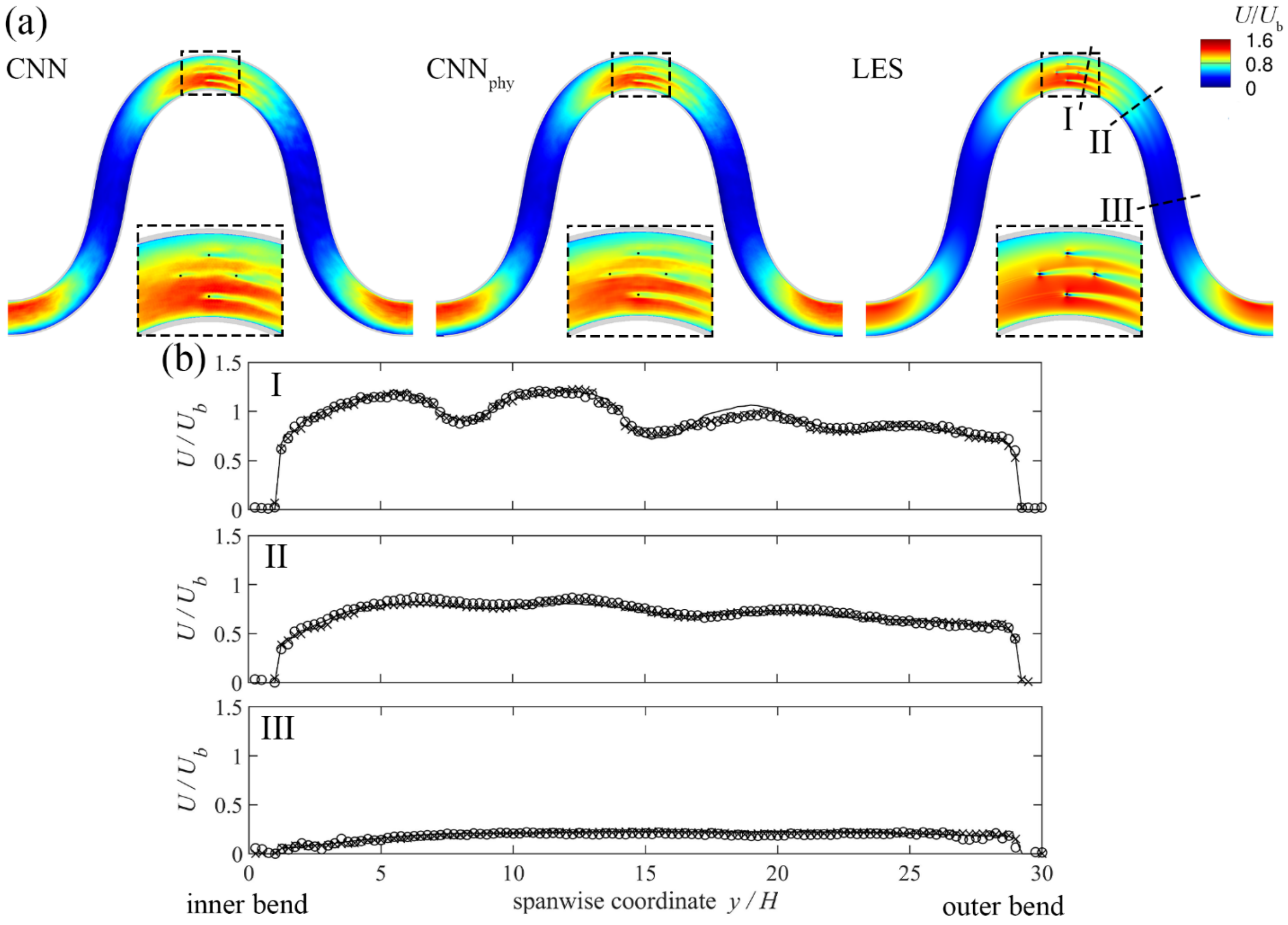

**Figure 17:** Time-averaged CNN predictions and LES results for the 3D flow field of the validation case. (a) Contours of streamwise velocity component ($\bar{u}$ / $U_b$) at the free surface of the virtual river from the top view. CNN represents the CNN model without physical constraint, $CNN_{phy}$ represents CNN model with physical constraint. In (a), flow is from left to right. (b) Profiles of the velocity component in the spanwise direction along the three dashed lines of I, II, and III in (a). In (b), solid lines represent the LES results, crosses represent CNN prediction with physical constraint, and circles represent CNN predictions without physical constraint.

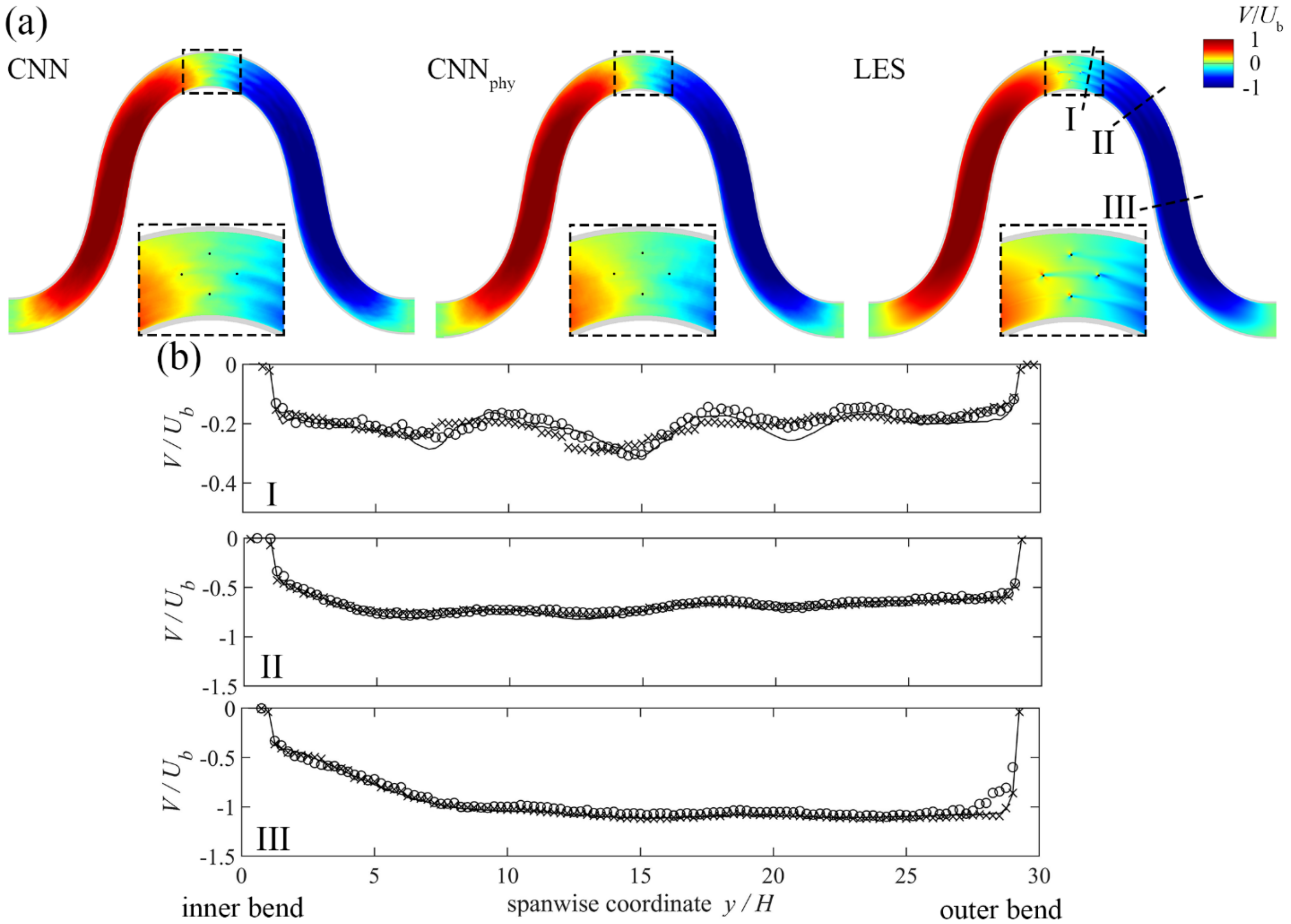

**Figure 18:** Time-averaged CNN predictions and LES results for the 3D flow field of the validation case. (a) Contours of spanwise velocity component ($\bar{v}/U_b$) at the free surface of the virtual river from the top view. CNN represents the CNN model without physical constraint, $CNN_{phy}$ represents CNN model with physical constraint. In (a), flow is from left to right. (b) Profiles of the velocity component in the spanwise direction along the three dashed lines of I, II, and III in (a). In (b), solid lines represent the LES results, crosses represent CNN prediction with divergence-free constraint, while circles represent CNN predictions without divergence constraint.

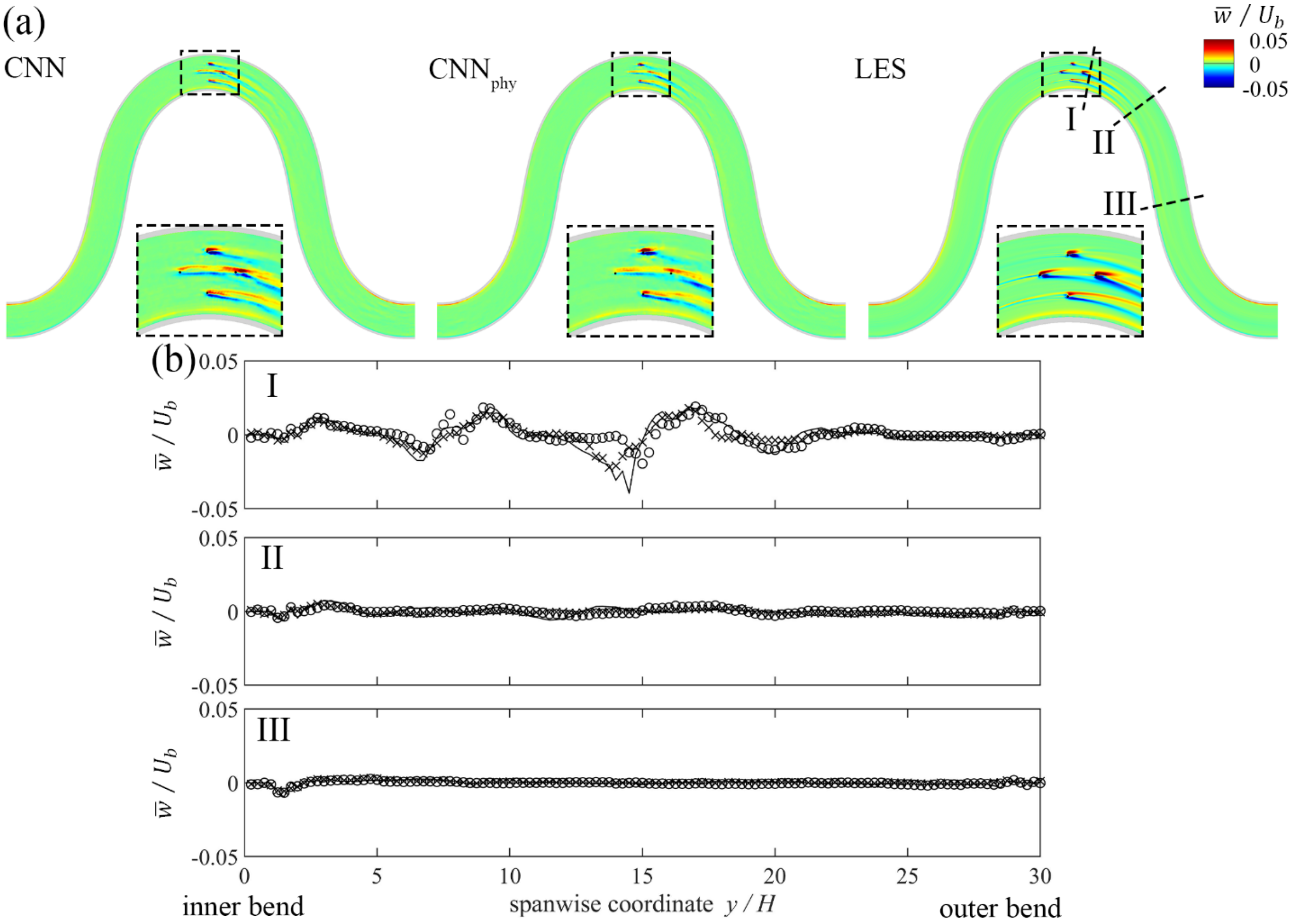

**Figure 19:** Time-averaged CNN predictions and LES results for the 3D flow field of the validation case. (a) Contours of vertical velocity component ($\overline{w}/U_b$) in the mid-depth of the virtual river from the top view. CNN represents the CNN model without physical constraint, and $CNN_{phy}$ represents CNN model with physical constraint. In (a), flow is from left to right. (b) Profiles of the velocity component in the spanwise direction along the three dashed lines of I, II, and III in (a). In (b), solid lines represent the LES results, crosses represent the CNN prediction with divergence-free constraint, and circles represent CNN predictions without divergence constraint.